\documentclass[prl,twocolumn,amsmath,amssymb,aps]{revtex4-2}

\usepackage{graphicx}
\usepackage{dcolumn}
\usepackage{bm}
\DeclareRobustCommand{\VAN}[3]{#2}
\let\VANthebibliography\thebibliography
\def\thebibliography{\DeclareRobustCommand{\VAN}[3]{##3}\VANthebibliography}
\usepackage[breaklinks,colorlinks = true,linkcolor = blue,urlcolor=blue,citecolor=blue]{hyperref}

\usepackage{graphicx}	
\usepackage{amsmath}	
\usepackage{amssymb}
\usepackage{nicefrac}

\begin{document}

\label{firstpage}
\title[Shock trapping and inertial escape]{Shock trapping and inertial escape: 
Dust-particle clustering in compressible turbulence}

\author{Anikat Kankaria}
\email{anikat.kankaria@gmail.com} 
\affiliation{International Centre for Theoretical Sciences, Tata Institute of Fundamental Research, Bangalore 560089, India}
\author{Samriddhi Sankar Ray}
\email{samriddhisankarray@gmail.com}
\affiliation{International Centre for Theoretical Sciences, Tata Institute of Fundamental Research, Bangalore 560089, India}

\begin{abstract}
We study the dynamics and clustering of dust particles with inertia in
	shock-dominated compressible turbulence using the two-dimensional,
	stochastically forced Burgers equation. At small Stokes numbers, shock
	trapping leads to extreme density inhomogeneities and nearly singular
	aggregation, with correlation dimensions approaching zero. With
	increasing inertia, particles undergo inertial escape and
	intermittently cross shock fronts, producing a sharp crossover from
	shock-dominated trapping to quasi-ballistic dynamics. This crossover is
	accompanied by a pronounced reduction in density fluctuations, a
	continuous increase of the correlation dimension from zero to the
	embedding dimension, and a power-law dependence of density fluctuations
	on the Stokes number over an extended intermediate regime. In this
	regime, particle distributions show scale-free coarse-grained density
	statistics arising from repeated trap--escape dynamics. This behaviour
	is qualitatively distinct from inertial-particle clustering in
	incompressible turbulence and is directly relevant to dust
	concentration in shock-rich regions of protoplanetary discs and other
	compressible astrophysical environments.
\end{abstract}
\maketitle



\section{Introduction}

The dynamics, spatial distribution, and transport of finite-inertia particles
suspended in turbulent flows lie at the heart of a wide range of natural and
astrophysical phenomena, from cloud microphysics~\citep[]{ShawReview2003,BodenshatzetalReview2010,BecGustavssonMehligReview2024},
star formation~\citep[]{FederrathKlessenReview2012}, and the evolution of protoplanetary discs~\citep[]{BirnstielFangJohansenReview2016,BirnstielReview2024}. 
A key challenge across all these systems is to understand how particles with non-zero Stokes
number interact with their carrier flow, and how different forms of
compressibility, ranging from weak density fluctuations to strongly
shock-dominated regimes, shape processes such as droplet coalescence, 
dust aggregation, and sedimentation~\citep{Mattssonetal2018}. While inertial-particle dynamics~\citep{Bec2003} and such issues are 
now relatively well understood in incompressible turbulence~\citep[]{Balkovskyetal2001,Falkovichetal2002,WilkinsonMehlig2003,WilkinsonMehlig2005,
Becetal2007,GustavssonMehlig2009,BecHomannRay2014,Sawetal2014,Becetal2016,JamesRay2017,Ray2018,Picardoetal2019}, 
similar questions in highly compressible,
shock-dominated turbulence are just beginning to be addressed~\citep{Yangetal2014,Xiaetal2016,Mattssonetal2018,Mattssonetal2019,Mattsson2020,Gerosaetal2023}. The statistical and
dynamical origins of preferential concentration~\citep{EatonFessler1994} in such 
flows are not yet well established, even though shocks and compressibility
strongly enhance intermittency and modify clustering behaviour in ways that
differ qualitatively from the incompressible case.

This gap in understanding is especially significant in the context of
protoplanetary discs. In these discs, inertial-particle clustering --- and the
resulting local dust density enhancements --- is a central ingredient in theories
of early planetesimal formation~\citep{Dominiketal2007,Johansenetal2007,Wilkinsonetal2008}. Despite decades of research, given the complexity of the problem, 
there is
little consensus on which mechanisms dominate dust concentration across the
vast range of disc environments~\citep[]{BirnstielFangJohansenReview2016,BirnstielReview2024}. Turbulent clustering~\citep{Cuzzietal2008},
streaming instabilities~\citep{YoudinGoodman2005}, vortex trapping~\citep{BargeSommeria1995,HengKenyon2010,LyraLin2013,Gibbonetal2015}, pressure bumps~\citep{Whipple1972,Pinilla2012}, gravitationally driven spirals~\citep{YoudinShu2002,Riceetal2004,Gibbonsetal2012}, and other processes may all contribute,
but none universally governs all disc regions.
Moreover, clustering is driven by rare, intermittent events that are
challenging to resolve in global three-dimensional (3D) simulations, and the underlying properties
of disc turbulence remain uncertain. 

Further, this challenge is compounded by the fact that realistic dust–gas simulations
must simultaneously capture drag, compressibility, shocks, shear,
stratification, pressure gradients, and possibly self-gravity --- a difficult  
combination to resolve properly at the resolutions needed to capture
intermittency. This motivates the use of idealised models that isolate specific physical mechanisms.
Among these, shock-dominated, compressible turbulence is
particularly relevant~\citep{Mattssonetal2018}: shocks are ubiquitous in discs, arising from spiral
density waves, zonal flows, gravitational instabilities, or magnetorotational 
turbulence. Yet their direct impact on dust concentration has been difficult to
quantify as global simulations entangle shocks with rotation and shear.

The stochastically forced two-dimensional (2D) Burgers
equation~\citep{Burgers1948}, provides a clean framework in which to study the
universal aspects of shock-driven particle
dynamics~\citep[]{FrischBecReview2001,BecKhaninReview2007}. Burgers
turbulence naturally generates a dynamically interacting network of shocks
whose statistics and intermittency can be measured with high accuracy.
Crucially, when the Burgers equation is driven by a power-law correlated
stochastic forcing, it develops an inertial-range energy spectrum close to the
Kolmogorov $k^{-5/3}$
scaling~\citep{ChekhlovYakhot1995,HayotJayaprakash1997,Mitraetal2005,Frischetal2013,DeMitraPandit2023,DeMitraPandit2024}.
This is a well-known property of stochastically forced Burgers turbulence
although the flow is strongly compressible and shock-dominated, the forcing
maintains a cascade that shares important features with real, compressible
astrophysical turbulence. This makes the model particularly attractive for
astrophysical applications because it captures both non-linear wave steepening
(and, therefore, shocks) and a realistic, turbulent-like
cascade~\citep[]{ShandarinZeldovichReview1989,MatarreseMohayaee2002,NeateTruman2007}. The
resulting environment produces inertial-particle dynamics --- shock trapping,
compression, intermittent expulsion --- that reflect general features absent in
incompressible flows but expected in many regions of protoplanetary discs.

The Burgers equation is a deliberately simplified model, lacking Keplerian
shear,  Coriolis forces, vertical stratification, pressure gradients, vortex
dynamics, and two-way drag coupling.  Thus, it should be viewed as a
proof-of-principle framework: a controlled setting in which the fundamental
effect of stochastic shocks on particle concentration can be extracted without
contamination from the full complexity of disc physics. By isolating these
effects, Burgers turbulence offers a clean test bed for identifying the
universal signatures of shock-driven clustering and for generating hypotheses
that can be tested with the shearing-box or global simulations.

In this work, we present an idealised study of inertial particles in
stochastically forced two-dimensional Burgers turbulence. We focus on
identifying and quantifying the fundamental mechanisms by which shocks
concentrate inertial particles, and on measuring how peak densities and aggregation structure,
scale with particle inertia. The stochastic forcing ensures both a persistent shock network and an
inertial-range cascade reminiscent of Kolmogorov turbulence, making the
resulting statistics relevant for astrophysical flows. Although the model omits
orbital and vertical dynamics, it provides robust predictions --- such as
Stokes-number scaling laws and concentration probability distributions --- that
illuminate whether transient shocks alone can generate the density enhancements
required for early planetesimal growth.

\section{Background and Numerical Details}

The two-dimensional, stochastically forced Burgers equation provides a minimal
yet dynamically rich model for compressible, shock-dominated flows and has
found increasing relevance in astrophysical contexts~\citep{ShandarinZeldovichReview1989,NeateTruman2007}. In protoplanetary discs,
molecular clouds, and early planetesimal-forming environments, turbulence is
often highly compressible, intermittency is extreme, and pressureless or
weak-pressure limits can be locally appropriate. In such regimes the Burgers
equation, with self-similar stochastic forcing, captures essential features such
as shock formation, filamentation, and strong density contrasts that govern
particle concentration and dust–gas interaction while retaining essential
elements of the spectral properties of turbulence. Its ability to generate
long-lived shocks and coherent structures makes it a useful surrogate for
studying clustering, caustics, and collision statistics of inertial particles
--- key mechanisms in early planet formation and dust coagulation. 

From a fluid-mechanics and statistical-physics perspective, stochastically
forced Burgers systems have long served as canonical models for turbulence
without vorticity, exhibiting exact shock solutions, non-Gaussian statistics,
anomalous scaling, and well-understood energy fluxes~\citep{BecKhaninReview2007,FrischBecReview2001}. They provide a clean
setting for studying intermittency, PDFs of velocity gradients, statistical
steady states under random forcing, and connections to the Kardar-Parisi-Zhang
(KPZ) universality class. As such, the 2D stochastically forced Burgers
equation sits at the intersection of astrophysical modeling and fundamental
studies of non-linear dynamics, offering analytic tractability alongside
strongly non-linear phenomenology.

\subsection{Governing Equations}

The compressible, shock dominated carrier flow ${\bf u}({\bf x},t)$, with a coefficient of kinematic viscosity $\nu$, 
is a solution of the two-dimensional Burgers equation 
\begin{equation}
	\frac{\partial {\bf u}}{\partial t} + ({\bf u} \cdot \nabla){\bf u} = \nu \nabla^2 {\bf u} + {\bf f}
\label{eq:burgers_eq}
\end{equation}
driven to a non-equilibrium statistically steady state by a zero mean, Gaussian, white-in-time, stochastic 
forcing with a spatial correlation 
\begin{equation}
   \langle \hat{f}(\mathbf{k},t) \cdot \hat{f}(\mathbf{k}', t') \rangle 
   = D_0 |\mathbf{k}|^{-2} \delta(\mathbf{k} + \mathbf{k}') \, \delta(t - t').
   \label{eq:force_covar}
\end{equation}
The forcing amplitude $D_0$ controls the energy injection rate and the specific choice of the covariance 
ensures that in steady state the kinetic energy spectrum scales as $k^{-5/3}$ in the inertial range~\citep{DeMitraPandit2023,DeMitraPandit2024}. 
Further, the choice of forcing and initial conditions ensure that the velocity field is vortex free.

We consider a dilute suspension of non-interacting dust grains, modeled as
small, heavy, inertial particles. Each particle, with position ${\bf x}_p$ and
velocity ${\bf v}_p$, is governed by the linear Stokes drag model~\citep{Bec2003}: 

\begin{eqnarray}
	\frac{d{\bf x}_p}{dt}  \,= \,{\bf v}_p(t);
   \label{eq:particle_pos} \\
	\frac{d {\bf v}_p}{dt} \, = \, -\frac{1}{\tau_p} \left[{\bf v}_p(t) - {\bf
	u}({\bf x}_p) \right].
   \label{eq:particle_vel}
\end{eqnarray}
The particle response time $\tau_p$ is non-dimensionalised by the characteristic, small-scale time-scale
$\tau_\eta = \sqrt{\nu / \epsilon}$, where $\epsilon = 2 \nu \langle |\nabla \mathbf{u}|^2 \rangle$ is 
the mean energy dissipation rate, to yield the Stokes number St = $\tau_p/\tau_\eta$.  

\begin{figure*}
\includegraphics[width=0.98\linewidth]{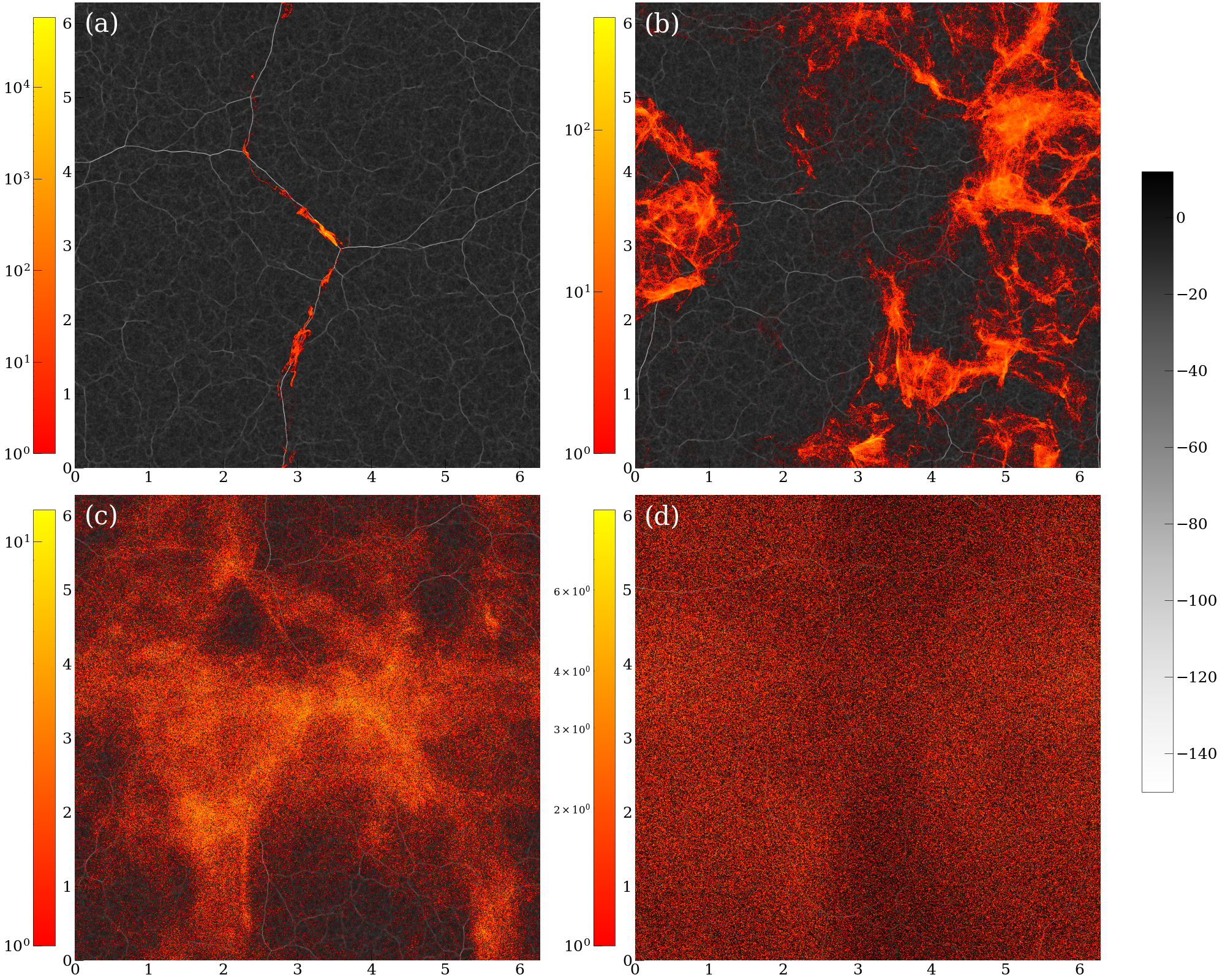}
	\caption{Representative snapshots of the coarse grained particle density fields $\Theta$ for (a) $St = 7.8$, (b) $St = 53.125$, 
	(c) $St = 500$, and (d) $St = 1250$, overlaid on the divergence $\nabla \cdot {\bf u}$ field (shown in grey scale) of the carrier flow. These snapshots are taken 
	after the non-equilibrium steady state has been reached for all the suspensions; an animation of the evolution of 
	the particles is given in Ref.~\citep{YT-movies}.}
\label{fig:densityfield}
\end{figure*}

\subsection{Direct Numerical Simulations}

We perform direct numerical simulations (DNSs) on a two-dimensional,
doubly-periodic domain of size $[0,2\pi]^2$. The equation~\eqref{eq:burgers_eq}
for the carrier flow is solved with a pseudo-spectral method on grids of up 
to $2048^2$ collocation points.  We impose the standard two-thirds truncation rule for dealiasing and a fourth-order integrating-factor
Runge–Kutta (IFRK4)~\citep{CoxMathews2002} scheme is used for time-marching. 

The stochastic force, with an ultra-violet cut-off to ensure that the
smallest-scale shocks mimic that of the unforced Burgers equation are resolved,
drives the system to nonequilibrium statistically steady state with an energy
spectrum which scales as $k^{-5/3}$ over an inertial range close to one and a
half decades in wavenumber space. 

The stochastic forcing~\eqref{eq:force_covar} injects energy in the available scales with a power-law,
and the system is evolved until a statistically stationary
nonequilibrium steady state (NESS) is achieved. Stationarity is confirmed by
the saturation of the total kinetic energy and the emergence of a persistent
power-law energy spectrum with a slope close to $-5/3$ over more than one and a
half decades in wavenumber~\citep{DeMitraPandit2023,DeMitraPandit2024}. 

An example of the network of (dynamically evolving) shocks can be seen in the background 
grey scale pseudo-color plots of $\nabla\cdot {\bf u}$ in Fig.~\ref{fig:densityfield}. The 
thin white curves denote the viscosity-broadened shocks which, as we shall see below, play 
a critical role in determining the fate of aggregates.

In this steady state we introduce $N_p = 2^{20} \approx 10^6$ particles for a
given Stokes number. We use a large range of Stokes numbers $0.1 \leq {\rm St}
\leq 3000$ and the particles are initially seeded randomly in the flow; their
subsequent evolution follow
Eqs.~\eqref{eq:particle_pos}-\eqref{eq:particle_vel} coupled with the
simultaneous solution of the carrier flow ${\bf u}$.  Typically, particle
positions are off-grid; hence we use a bilinear interpolation scheme to obtain
the fluid velocity ${\bf u}({\bf x}_p)$ at their locations. We now evolve this
mixed phase suspension over several large-eddy turnover times and ensure, as we
shall see below, that the particles converge to their stationary distribution.

\section{Shock-induced Inhomogeneity in Particle Number Density} 

We begin with a simple question: how does the stationary distribution of particles look like and what role 
does the Stokes number play? This is most conveniently seen through the particle number density 
$\Theta ({\bf x},t) \equiv \nicefrac{\#_p({\bf x},t)}{N_p}$~\citep{Yangetal2014}, 
defined at position ${\bf x}$ and time $t$, where $\#_p({\bf x},t)$ is 
the number of particles in a square, centered at ${\bf x}$ with sides $\ell = \nicefrac{4\pi^2}{N_p}$. 
In the vanishing Stokes limit, and unlike the homogeneous distribution of particles in incompressible 
flows, we expect an extremely inhomogeneous distribution of $\Theta$ as most particles converge on the filamentary 
network of shock lines. In sharp contrast, in the large Stokes limit, the particles ought to decouple from the 
underlying flow and distribute uniformly in a manner similar to what is seen in incompressible turbulence. 

In Fig.~\ref{fig:densityfield}, we show plots of $\Theta$, for different Stokes
numbers, at long times when the evolution of particles have converged to a
steady state distribution. In particular, in the small Stokes limit
(Fig.~\ref{fig:densityfield}(a), St = 7.8), the pseudocolor plot of the number
density shows a spike on a subset of the shocks while being close to 0
elsewhere which is consistent with the physical intuition developed earlier. Likewise,
in the large Stokes limit, as seen in Fig.~\ref{fig:densityfield}(d) for St =
1250,  $\Theta$ shows a near uniform distribution. In between such Stokes
numbers , as seen in panels (b) and (c) of Fig.~\ref{fig:densityfield}, the
density field ought to transition from the void-dominated, preferentially
concentrated $St \to 0$ limit to the space-filling large Stokes asymptotics.
This seemingly monotonic transition, in Stokes number, from an inhomogeneous to
a homogeneous particle distribution in shock dominated flows is markedly
different from what one sees in incompressible turbulence. In incompressible
flows particles distribute uniformly in both St $\to 0$ and St $\to \infty$
limits.

\begin{figure}
\includegraphics[width=1.0\linewidth]{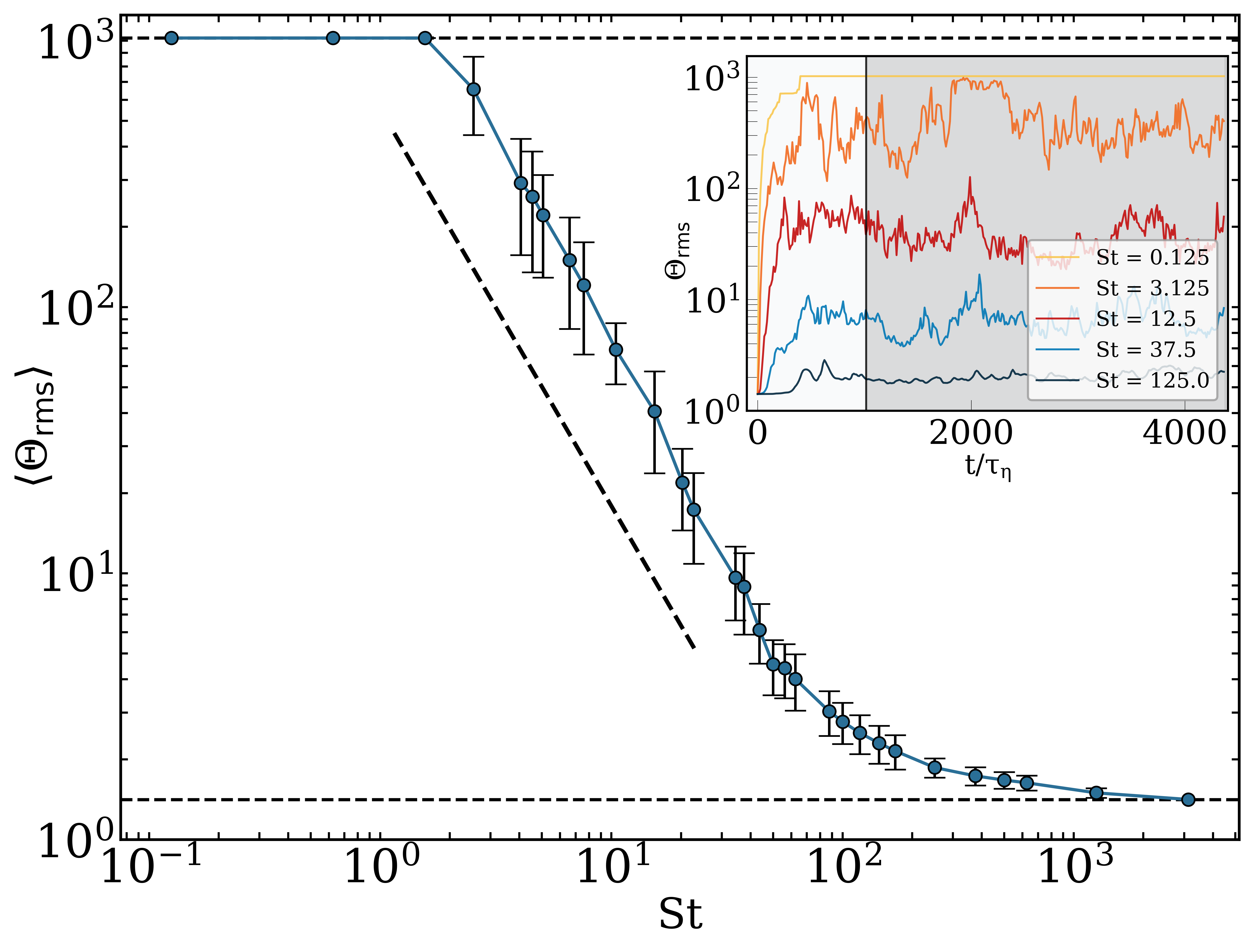}
	\caption{(Inset) Plots of the evolution of $\Theta_{\rm rms}$ --- measure of the inhomogeneity in particle distribution (see text) --- 
	as a function of the non-dimensional time $\tau = t/\tau_{\rm \eta}$ 
	for representative values of the Stokes number (see legend). The shaded region corresponds to the steady state distribution of particles from whence 
	we calculate the mean value $\langle \Theta_{\rm rms} \rangle$ of $\Theta_{\rm rms}$ over this steady state for different $St$ numbers. In the main panel 
	we plot $\langle \Theta_{\rm rms} \rangle$ vs $St$. The error bars on these are the standard deviation in values of $\Theta_{\rm rms}$ over the 
	stationary regime (see inset). The two, dashed horizontal lines denote two different limiting behaviours. The upper one 
	$\langle \Theta_{\rm rms} \rangle  = \sqrt{N_p}$ corresponds to all particles collapsing to a singular point whereas
	the lower one $\langle \Theta_{\rm rms} \rangle  = \sqrt{2}$ emerges as the asymptotic limit for a uniform distribution with Poisson fluctuations (see 
	text).}
\label{fig:thetarms}
\end{figure}

The particle number density field $\Theta$ is a convenient tool to quantify the extent of clustering for a given 
Stokes number. However,  by definition because of the normalisation factor 
$\nicefrac{1}{N_p}$, its mean value $\langle \Theta (t) \rangle = \int_{[0,2\pi]^2} 
\Theta({\bf x},t)\, d{\bf x}$ 
is unity at all times and for all Stokes numbers. Hence, we choose a different measure $\Theta_{\rm rms}(t) = \sqrt{\nicefrac{1}{N_p}\int_{[0,2\pi]^2}\Theta(\mathbf{x},t)^2}$. 
Such a definition allows a variation in $\Theta_{\rm rms}$ --- unlike what we would have for  $\langle \Theta \rangle$ --- 
as the degree of density concentration varies with changing Stokes number. In the inset of Fig.~\ref{fig:thetarms}, 
we show semilog plots of $\Theta_{\rm rms}$ versus the non-dimensional time $t/\tau_\eta$ for a few representative values of the 
Stokes number (see legend). After an initial transient, as the density field concentrates from an initial uniform 
distribution, $\Theta_{\rm rms}$ converges to a steady state value over time indicated by the shaded region in the plot.
In this shaded region, we are able to calculate the mean value of $\Theta_{\rm rms}$ which gives a ``concentration parameter'' 
$\langle \Theta_{\rm rms} \rangle$; the standard deviation of $\Theta_{\rm rms}$ over the steady state gives us a measure of the 
statistical error on this concentration parameter.

In the main panel of Fig.~\ref{fig:thetarms} we plot the $\langle \Theta_{\rm
rms} \rangle$ versus the Stokes number on a semilog plot.  Several things stand
out. In the tracer limit, by definition $\langle \Theta_{\rm rms} \rangle \to
\sqrt{N_p}$; the upper horizontal, dashed line indicates this value and we find
our results from DNSs converge to this limit. In the large Stokes limit, we
find $\langle \Theta_{\rm rms} \rangle \to \sqrt{2}$ as indicated by the lower,
horizontal dashed limit. This $\sqrt{2}$ limit is a consequence of a random, but homogeneous, 
distribution of $N_p$ particles: this is a Poisson process and it follows, trivially, that 
for such distributions $\langle \Theta_{\rm rms} \rangle = \sqrt{2}$. Indeed, our initial 
random distribution of particles also show a $\sqrt{2}$ measure~\citep{Yangetal2014}.

Between these two asymptotics, in the  range $\mathcal{O}(1) \lesssim {\rm St}
\lesssim \mathcal{O}(100)$,  we see the transition from the regime dominated by
shock-trapped-particles to one in which inertial effects matter. In this
reasonably narrow range of Stokes numbers, $\langle \Theta_{\rm rms} \rangle$
decreases rapidly as the dynamics of the dusty flow alters significantly. Our
numerical data suggests (see Fig.~\ref{fig:thetarms}) that the transition
between the $\sqrt{N_p}$ and $\sqrt{2}$ limits is self-similar: indeed, it
seems to follow a power-law with $\langle \Theta_{\rm rms} \rangle \sim {\rm
St}^{-3/2}$ in the range $\mathcal{O}(1) \lesssim {\rm St} \lesssim
\mathcal{O}(100)$ (see Fig.~\ref{fig:thetarms}) where the  particle-density field results from a subtle
interplay between the particle-inertia on one hand and the shocks on the other. 

This subtle interplay reflects in the error-bars on the measured $\langle
\Theta_{\rm rms} \rangle$.  While the particle dynamics --- and hence the
density field --- are completely dictated by either shocks (St $\ll 1$) or particle 
inertia (St $\gg 1$), in the transition regime $\mathcal{O}(1) \lesssim {\rm
St} \lesssim \mathcal{O}(100)$ particles possess sufficient inertia to
intermittently escape from shocks and yet remain susceptible to recapture as
shocks form, coalesce, and annihilate. Hence, the larger fluctuations in
$\Theta_{\rm rms}$, seen in the inset of Fig.~\ref{fig:thetarms}, and reflected
in the larger error-bars on the measured $\langle \Theta_{\rm rms} \rangle$. 

\section{Density field Inhomogeneity at the Mesoscale} 

\begin{figure}
\includegraphics[width=1.0\linewidth]{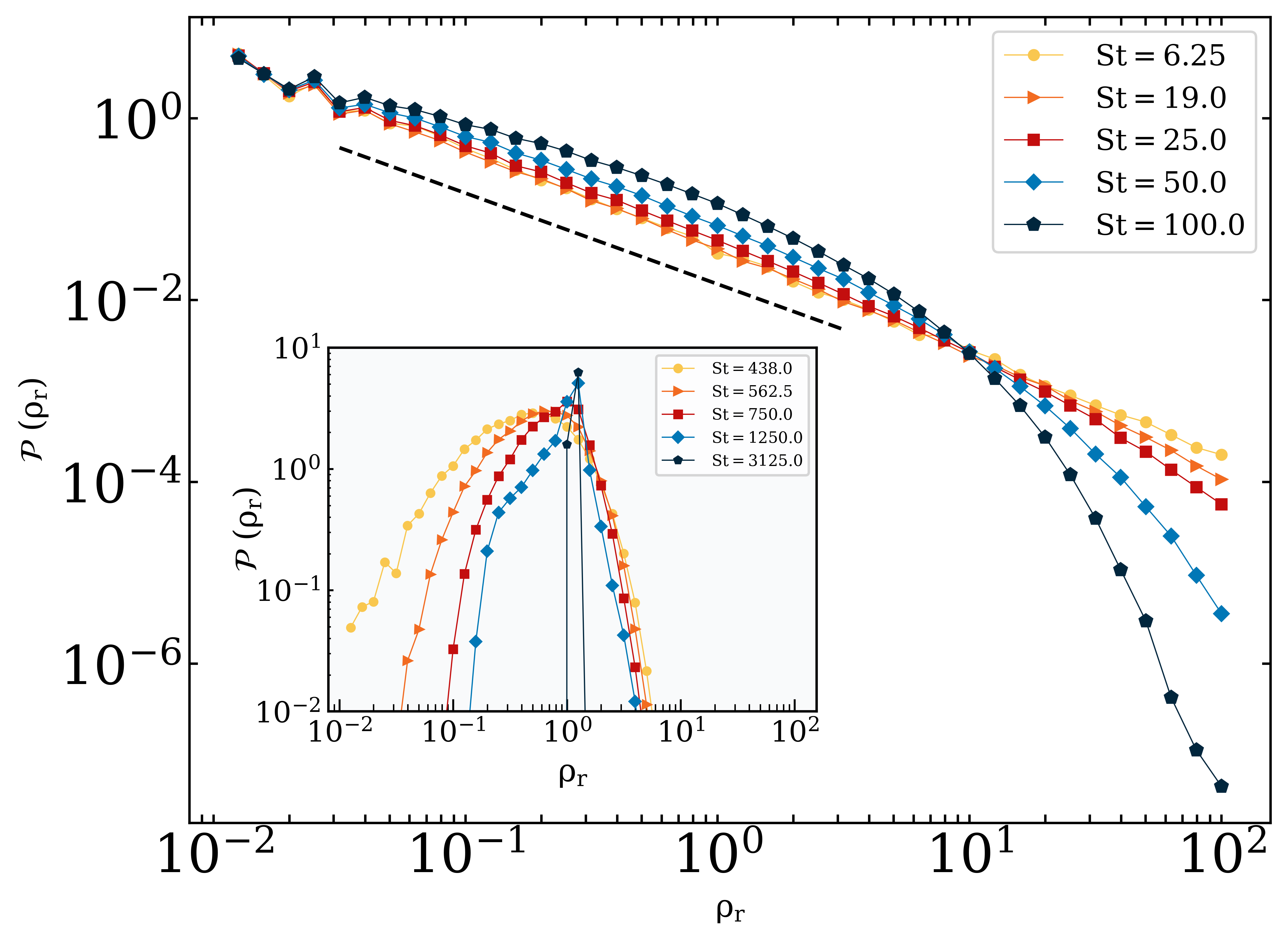}
	\caption{Loglog plots of the probability density functions of the particle density fields coarse grained at scales 
	${r} = 0.15$, for different Stokes numbers (see legend).}
\label{fig:QLpdf}
\end{figure}

The mean particle number density $\Theta$ captures fluctuations at the scale of the particles by definition. However, as 
evident in Fig.~\ref{fig:densityfield}, inhomogeneity also show up at mesoscales well beyond the particle scale $\nicefrac{4\pi^2}{N_p}$.
Such ideas of a scale-dependent aggregation are familiar to those working on turbulent transport in incompressible flows.
We adapt a similar prescription to define a coarse-grained density 
\begin{equation}
\rho_r(\mathbf{x}) = \frac{\#(\mathbf{x}) 4\pi^2}{N_p r^2}
\end{equation}
from non-overlapping squares of side $r$ (larger than the smallest length scales of the flow), 
centered at ${\bf x}$ with $\#$ being the 
number of particles in such squares  normalized by the mean density $\nicefrac{N_p}{4\pi^2}$.
The probability distribution $\mathcal{P}({\rho_r})$ of this coarse-grained density gives a measure 
of the nature of voids and cluster over the mesoscale $\eta \ll r \ll 2\pi$.

The nature of this distribution $\mathcal{P}({\rho_r})$, in the small St $\ll 1$
and large St $\gg 100$ limits follow from the observations in
Fig.~\ref{fig:densityfield}; while for vanishing Stokes numbers, the
(non-interacting) particles collapse on to a near singular point, at large
Stokes number the random, homogeneous distribution ought to lead to a Poisson
distribution as seen in incompressible turbulence.  The non-trivial result
comes about for intermediate Stokes numbers $10 \lesssim {\rm St} \lesssim 200$
for which, as seen earlier, $\sqrt{2} \ll \langle \Theta_{\rm rms} \rangle \ll
\sqrt{N_p}$ lie between the two limiting cases.

For such Stokes numbers, the particle dynamics suggest
(Fig.~\ref{fig:densityfield}(b)-(c) and more clearly in the animation of the
time-evolution of particles~\citep{YT-movies}) a continuing struggle between
inertia-induced escape and shock-induced trapping. This leads to typical
arrangement of particles on filamentary structures with a fluctuating particle
density per unit length $\mu$ and, hence, at any instance of time for a square of side $r$, the number of 
particles $\# \sim \mu r$ with the result $\rho_r \sim \nicefrac{4\pi^2 \mu}{N_p r} \sim \mu$.
Therefore, the probability density $\mathcal{P}({\rho_r})$ must be rooted in the probability density function 
$\mathcal{P}({\mu})$ of the (fluctuating) particle density $\mu$. 

This brings us to the question of what could be the simplest model for $\mu$. We know that as particles evolve in
time, $\mu$ increases from the effects of the shocks to trap and decreases as
inertia leads to a more spreading out of such particles. This continuous game
of traps and escapes suggests a simple, relation between the values of $\mu$ at
discrete ($n$ and $n+1$) intervals: $\mu_{n+1} = \alpha_n \mu_n$.  Here,
$\alpha_n$ is a random positive variable which can pick values greater or lesser than 1,
depending on whether the local particle density increases (shock-induced
trapping) or decreases (inertia-induced escapes). This random, multiplicative
\textit{cascade} picture, to a first approximation when the excursions of
$\alpha_n$ are not very large (as is clear from our observations of the
particle dynamics), suggests an approximately constant distribution $\mathcal{P}({\log \mu})$ of the log
of $\mu$. It is trivial to show that this implies $\mathcal{P}({\mu}) \sim
\mu^{-1}$ and hence $\mathcal{P}({\rho_r}) \sim \rho_r^{-1}$.

Thus, a na\"ive theoretical argument for intermediate Stokes numbers $10
\lesssim {\rm St} \lesssim 200$ based on the observation that particles
alternately collapse onto shock structures and subsequently smear out on the
filamentary structures of the system spanning shocks, producing a
(multiplicative) sequence of compressions and expansions results in scale-free
fluctuations in the coarse-grained density field with a precise inverse
power-law form for $\mathcal{P}({\rho_r})$.

\section{The Structure of Aggregates} 

\begin{figure*}
\includegraphics[width=0.48\linewidth]{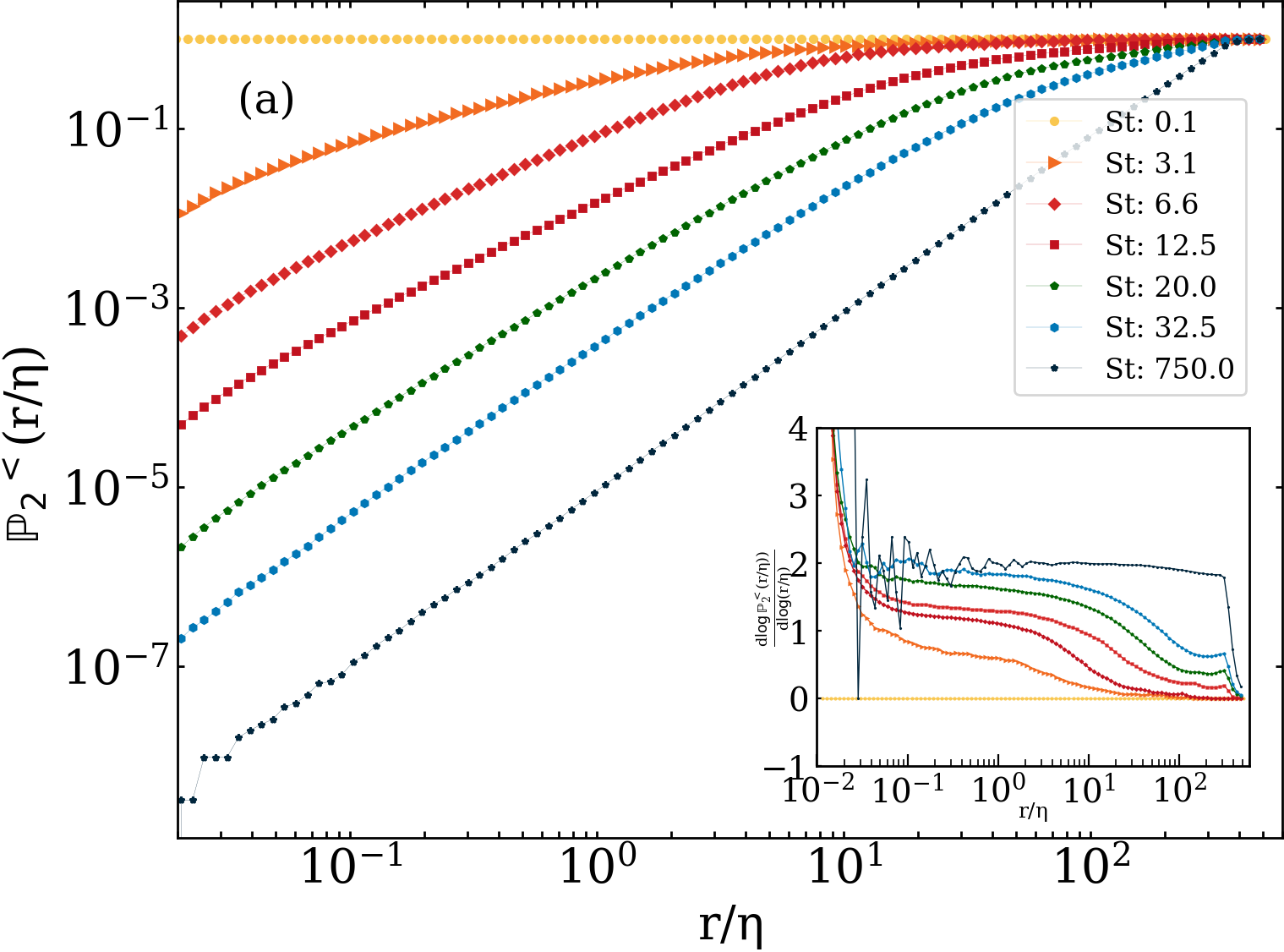}
\includegraphics[width=0.462\linewidth]{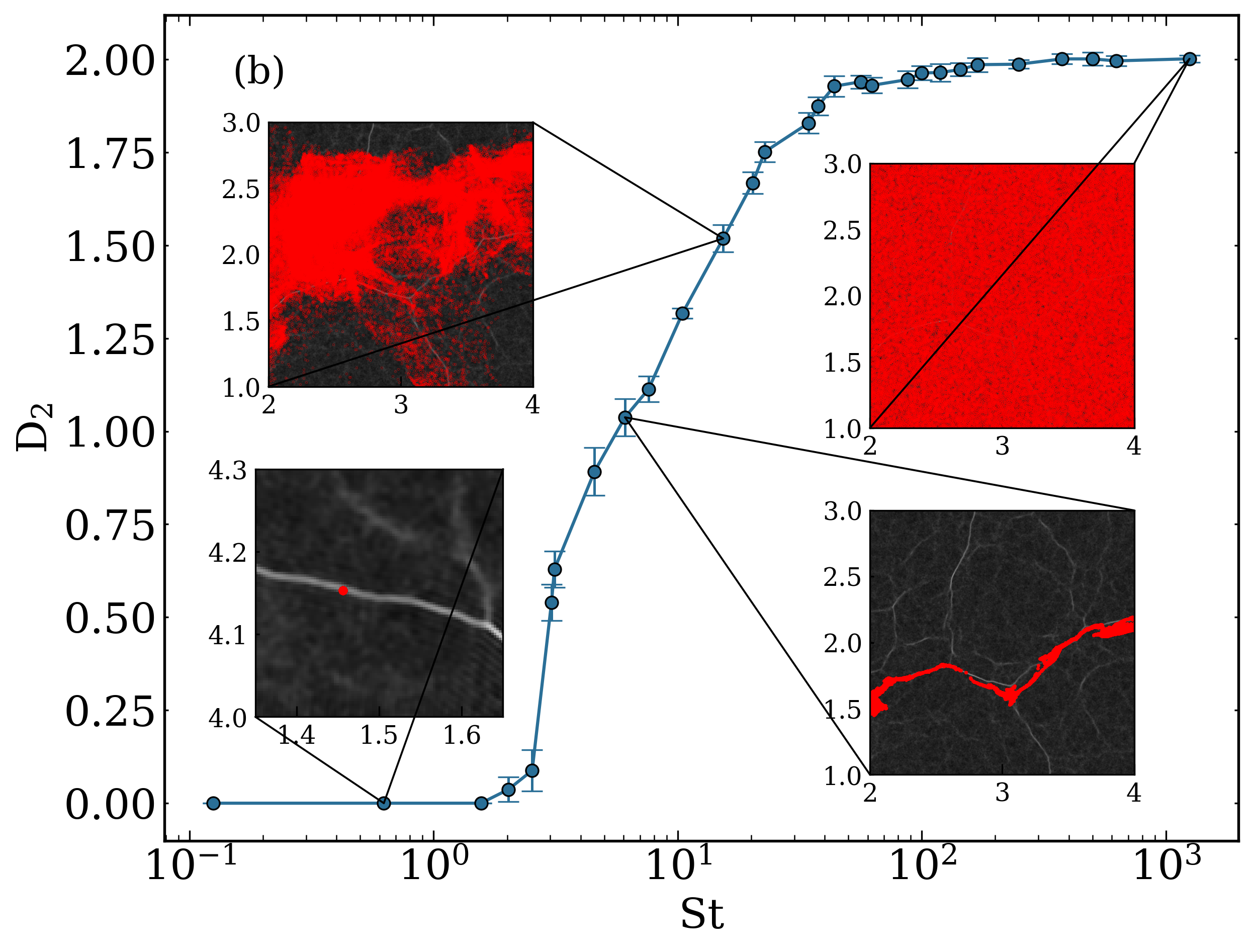}
	\caption{(a) Loglog plots of the cumulative density function $P^< (r)$ vs $r/\eta$ for different Stokes numbers showing a clear power-law 
scaling. From a local slope analysis of these power laws we estimate the correlation dimension $D_2$ (and its error bars) which are shown, as a function 
	of $St$, in panel (b). In panel (b) we also include several insets showing the particle positions which illustrate the corresponding 
	fractal dimension $D_2$ of their areas.}
\label{fig:D2}
\end{figure*}

Our measurements of $\langle \Theta_{\rm rms} \rangle$ as seen in
Fig.~\ref{fig:thetarms}, and confirmed from snapshots of $\Theta$ as shown in
Fig.~\ref{fig:densityfield}, indicate precisely the inhomogeneous nature of the
density field. However, they do not inform us of the structure of the aggregates
especially for Stokes numbers where $\langle \Theta_{\rm rms} \rangle \gg
\sqrt{2}$. In particular, the fractality of these aggregates at small-scales are yet to be quantified.

The time honored way to deal with this question is through the measurement of
the probability $\mathbb{P}^<_2$ of finding a pair of particles within a
distance $r/\eta$; the normalisation by $\eta$ ensures that this is
non-dimensional. If aggregates form (fractal) clusters, this
pair-correlation probability would be self-similar $\mathbb{P}^<_2 \sim
(r/\eta)^{D_2}$ with $D_2$ being known as the correlation dimension.  Therefore,
the variation of $D_2$ with the Stokes number provides a direct measure of the
spatial organization of particles --- the structure of the aggregates
--- with increasing inertia. 

We see clear evidence of this self-similar
behaviour of $\mathbb{P}^<_2$ in Fig.~\ref{fig:D2}(a) for a few representative
Stokes numbers.  From local slopes, such as the one shown in the inset of
Fig.~\ref{fig:D2}(a), one can estimate $D_2$ as the mean value, and the error
bars on the exponents as the standard deviation, of $\frac{d\log
\mathbb{P}^<_2}{d \log r/\eta}$ in the range of $r/\eta$ where a power-law
exists.

As we have seen before in Fig.~\ref{fig:densityfield}(a), as well as in our
measurement of $\langle \Theta_{\rm rms} \rangle$, for $St \lesssim 1$
particles exhibit strong localization and collapse into nearly singular
concentration as they get trapped in shocks. As the flow evolves, the
continual shock merging and multiplying drive this aggregation into highly
concentrated regions which eventually collapse onto nearly singular,
zero-dimensional points and stay ``stuck''. Our measurement of $D_2$ for such
small Stokes number, as seen in Fig.~\ref{fig:D2}(b), confirm this as we see
$D_2 \to 0$ as St $\to 0$. This corresponds to effectively
0-dimensional ($D_2 = 0$) aggregates as shown in the snapshot from our data
corresponding to St = 0.6 in  Fig.~\ref{fig:D2}(b).

In the large Stokes limit $St \sim \mathcal{O}(100)$ (as suggested in our
measurements of $\langle \Theta_{\rm rms} \rangle$), the particle inertia is
sufficiently large such that shocks no longer impede their motion. In this
regime, particles decouple from the compressive structures and move
quasi-ballistically, leading to a saturation of the correlation dimension to
the embedding dimension of the flow: $\mathcal{D}_2 \rightarrow 2$ and, hence, 
a nearly homogeneous spatial distribution of particles. In Fig.~\ref{fig:D2}(b), 
we see this to be the case in our measurement and, for visual clarity, the inset 
showing a snapshot for St = 1250 clearly shows a homogeneous, space-filling $D_2 = d = 2$ 
cluster.

Like in our measurements of $\langle \Theta_{\rm rms} \rangle$, the interesting observation 
is the transition between these two limits. For instance, in Fig.~\ref{fig:D2}(b) we find a sharp 
transition as we go from $D_2 \to 0$ to $D_2 \to 2$ which crosses the $D_2 = 1$ around 
St = 10. The correlation dimension being unity is suggestive of particles that 
have sufficient inertia to avoid the singular collapse seen for St $\ll 1$ but not enough to 
avoid being largely concentrated along the unit-dimensional, filamentary network of shocks as suggested 
already in Fig.~\ref{fig:densityfield}(a).  This phenomenon is confirmed when we look at a snapshot 
of particles, in the inset accompanying the $D_2 \approx 1$ point in Fig~\ref{fig:D2}(b). Furthermore, 
for slightly larger Stokes numbers, similar snapshots of particle position 
give a visual cue of aggregates whose dimensions must lie between the filamentary one-dimensional structures 
and the space-filling two-dimensional structures as seen in in Fig~\ref{fig:D2}(b) for St = 30 with 
$D_2 = 1.6$.

\section{Conclusions}

We have investigated the dynamics and spatial organisation of inertial
particles in a stochastically forced, shock-dominated Burgers flow, with the
aim of isolating the role of compressive turbulence in particle concentration.
This deliberately minimal framework allows us to identify the dynamical
consequences of shocks in a controlled setting, independent of rotation, shear,
stratification, or self-gravity, while retaining key features of strongly
compressible astrophysical turbulence.

In the small-Stokes regime ($\mathrm{St} \ll 1$), particles are tightly coupled
to the gas and rapidly converge into shock-compressed regions. Shocks act as
efficient dynamical traps, leading to extreme density enhancements and nearly
singular clustering, as reflected by large coarse-grained density fluctuations
and correlation dimensions approaching zero. The resulting long residence times
in compressive regions suggest that shock trapping may provide a simple and
robust pathway for enhanced grain–grain encounters even for particles with very
small inertia.

At intermediate Stokes numbers ($\mathrm{St} = \mathcal{O}(1)$), particle
response times become comparable to characteristic shock and strain timescales.
Particles intermittently detach from and reattach to shocks as the compressive
network evolves, producing the strongest intermittency in space and time.
Aggregates transition from point-like clusters to filamentary structures
aligned with shocks, accompanied by scale-free coarse-grained density
statistics over inertial-range scales.

For large inertia ($\mathrm{St} \gg 1$), particles increasingly decouple from
the carrier flow and cross shocks quasi-ballistically. Density fluctuations
weaken, the correlation dimension approaches the embedding dimension, and the
spatial distribution becomes nearly homogeneous, consistent with shock-induced
concentration becoming inefficient.

It is noteworthy that, despite the absence of self-gravity, shock-driven
trapping at small Stokes numbers leads to near-singular aggregation and
collapse-like behaviour~\citep{GoldhirschReview2003,Becetal2013}. In this sense, the dynamics provide a kinematic,
gravity-free analogue of gravitational collapse, in which dissipation and
compressive shocks act as an effective focusing mechanism. Unlike true
gravitational collapse, however, the aggregation here is entirely driven by the
externally forced flow and does not involve long-range attraction or feedback
from the particle distribution.

Finally, we note that in realistic disc environments the relevant Stokes number
is defined relative to the local orbital frequency rather than the smallest
turbulent timescale, and that the mapping between the two depends on the nature
of the turbulence. With this caveat, our results nonetheless demonstrate that
shocks alone can act as a robust driver of particle clustering, establishing a
clean baseline for understanding shock-mediated trapping and escape in
compressible astrophysical flows. It remains to be seen in future work how this
trap-escape mechanism affects longer, \textit{stringier} particles whose length
scales are comparable to the domain-spanning network of
shocks~\citep{Picardoetal2020}, anisotropic particles with additional 
degrees of freedom~\citep{Anandetal2020} and the effect of a two-way coupled dust-gas 
interaction~\citep[]{Pandeyetal2019,JoshiRay2025}.

\section*{Acknowledgements}

AK would like to acknowledge Rajarshi Chattopadhyay, Sandip Sahoo, and Anwesha
Dey.  SSR acknowledges  the Indo–French Centre for the Promotion of Advanced
Scientific Research (IFCPAR/CEFIPRA, project no. 6704-1) for support.  This
research was supported in part by the International Centre for Theoretical
Sciences (ICTS) for the program --- 10th Indian Statistical Physics Community
Meeting (code: ICTS/10thISPCM2025/04). The simulations were performed on the
ICTS clusters Mario, Tetris, and Contra.  AK also acknowledges Google
Colaboratory environment where the data generation and analysis were performed.
AK and SSR acknowledge the support of the DAE, Government of India, under
projects nos.  12-R\&D-TFR-5.10-1100 and RTI4001.




\bibliography{references}

\begin{thebibliography}{60}%
\makeatletter
\providecommand \@ifxundefined [1]{%
 \@ifx{#1\undefined}
}%
\providecommand \@ifnum [1]{%
 \ifnum #1\expandafter \@firstoftwo
 \else \expandafter \@secondoftwo
 \fi
}%
\providecommand \@ifx [1]{%
 \ifx #1\expandafter \@firstoftwo
 \else \expandafter \@secondoftwo
 \fi
}%
\providecommand \natexlab [1]{#1}%
\providecommand \enquote  [1]{``#1''}%
\providecommand \bibnamefont  [1]{#1}%
\providecommand \bibfnamefont [1]{#1}%
\providecommand \citenamefont [1]{#1}%
\providecommand \href@noop [0]{\@secondoftwo}%
\providecommand \href [0]{\begingroup \@sanitize@url \@href}%
\providecommand \@href[1]{\@@startlink{#1}\@@href}%
\providecommand \@@href[1]{\endgroup#1\@@endlink}%
\providecommand \@sanitize@url [0]{\catcode `\\12\catcode `\$12\catcode
  `\&12\catcode `\#12\catcode `\^12\catcode `\_12\catcode `\%12\relax}%
\providecommand \@@startlink[1]{}%
\providecommand \@@endlink[0]{}%
\providecommand \url  [0]{\begingroup\@sanitize@url \@url }%
\providecommand \@url [1]{\endgroup\@href {#1}{\urlprefix }}%
\providecommand \urlprefix  [0]{URL }%
\providecommand \Eprint [0]{\href }%
\providecommand \doibase [0]{https://doi.org/}%
\providecommand \selectlanguage [0]{\@gobble}%
\providecommand \bibinfo  [0]{\@secondoftwo}%
\providecommand \bibfield  [0]{\@secondoftwo}%
\providecommand \translation [1]{[#1]}%
\providecommand \BibitemOpen [0]{}%
\providecommand \bibitemStop [0]{}%
\providecommand \bibitemNoStop [0]{.\EOS\space}%
\providecommand \EOS [0]{\spacefactor3000\relax}%
\providecommand \BibitemShut  [1]{\csname bibitem#1\endcsname}%
\let\auto@bib@innerbib\@empty
\bibitem [{\citenamefont {Shaw}(2003)}]{ShawReview2003}%
  \BibitemOpen
  \bibfield  {author} {\bibinfo {author} {\bibfnamefont {R.~A.}\ \bibnamefont
  {Shaw}},\ }\bibfield  {title} {\bibinfo {title} {Particle-turbulence
  interactions in atmospheric clouds},\ }\href
  {https://doi.org/https://doi.org/10.1146/annurev.fluid.35.101101.161125}
  {\bibfield  {journal} {\bibinfo  {journal} {Annual Review of Fluid
  Mechanics}\ }\textbf {\bibinfo {volume} {35}},\ \bibinfo {pages} {183}
  (\bibinfo {year} {2003})}\BibitemShut {NoStop}%
\bibitem [{\citenamefont {Bodenschatz}\ \emph {et~al.}(2010)\citenamefont
  {Bodenschatz}, \citenamefont {Malinowski}, \citenamefont {Shaw},\ and\
  \citenamefont {Stratmann}}]{BodenshatzetalReview2010}%
  \BibitemOpen
  \bibfield  {author} {\bibinfo {author} {\bibfnamefont {E.}~\bibnamefont
  {Bodenschatz}}, \bibinfo {author} {\bibfnamefont {S.~P.}\ \bibnamefont
  {Malinowski}}, \bibinfo {author} {\bibfnamefont {R.~A.}\ \bibnamefont
  {Shaw}},\ and\ \bibinfo {author} {\bibfnamefont {F.}~\bibnamefont
  {Stratmann}},\ }\bibfield  {title} {\bibinfo {title} {Can we understand
  clouds without turbulence?},\ }\href
  {https://doi.org/10.1126/science.1185138} {\bibfield  {journal} {\bibinfo
  {journal} {Science}\ }\textbf {\bibinfo {volume} {327}},\ \bibinfo {pages}
  {970} (\bibinfo {year} {2010})},\ \Eprint
  {https://arxiv.org/abs/https://www.science.org/doi/pdf/10.1126/science.1185138}
  {https://www.science.org/doi/pdf/10.1126/science.1185138} \BibitemShut
  {NoStop}%
\bibitem [{\citenamefont {Bec}\ \emph {et~al.}(2024)\citenamefont {Bec},
  \citenamefont {Gustavsson},\ and\ \citenamefont
  {Mehlig}}]{BecGustavssonMehligReview2024}%
  \BibitemOpen
  \bibfield  {author} {\bibinfo {author} {\bibfnamefont {J.}~\bibnamefont
  {Bec}}, \bibinfo {author} {\bibfnamefont {K.}~\bibnamefont {Gustavsson}},\
  and\ \bibinfo {author} {\bibfnamefont {B.}~\bibnamefont {Mehlig}},\
  }\bibfield  {title} {\bibinfo {title} {Statistical models for the dynamics of
  heavy particles in turbulence},\ }\href
  {https://doi.org/https://doi.org/10.1146/annurev-fluid-032822-014140}
  {\bibfield  {journal} {\bibinfo  {journal} {Annual Review of Fluid
  Mechanics}\ }\textbf {\bibinfo {volume} {56}},\ \bibinfo {pages} {189}
  (\bibinfo {year} {2024})}\BibitemShut {NoStop}%
\bibitem [{\citenamefont {Federrath}\ and\ \citenamefont
  {Klessen}(2012)}]{FederrathKlessenReview2012}%
  \BibitemOpen
  \bibfield  {author} {\bibinfo {author} {\bibfnamefont {C.}~\bibnamefont
  {Federrath}}\ and\ \bibinfo {author} {\bibfnamefont {R.~S.}\ \bibnamefont
  {Klessen}},\ }\bibfield  {title} {\bibinfo {title} {The star formation rate
  of turbulent magnetized clouds: Comparing theory, simulations, and
  observations},\ }\href {https://doi.org/10.1088/0004-637X/761/2/156}
  {\bibfield  {journal} {\bibinfo  {journal} {The Astrophysical Journal}\
  }\textbf {\bibinfo {volume} {761}},\ \bibinfo {pages} {156} (\bibinfo {year}
  {2012})}\BibitemShut {NoStop}%
\bibitem [{\citenamefont {Birnstiel}\ \emph {et~al.}(2016)\citenamefont
  {Birnstiel}, \citenamefont {Fang},\ and\ \citenamefont
  {Johansen}}]{BirnstielFangJohansenReview2016}%
  \BibitemOpen
  \bibfield  {author} {\bibinfo {author} {\bibfnamefont {T.}~\bibnamefont
  {Birnstiel}}, \bibinfo {author} {\bibfnamefont {M.}~\bibnamefont {Fang}},\
  and\ \bibinfo {author} {\bibfnamefont {A.}~\bibnamefont {Johansen}},\
  }\bibfield  {title} {\bibinfo {title} {Dust evolution and the formation of
  planetesimals},\ }\href {https://doi.org/10.1007/s11214-016-0256-1}
  {\bibfield  {journal} {\bibinfo  {journal} {Space Science Reviews}\ }\textbf
  {\bibinfo {volume} {205}},\ \bibinfo {pages} {41} (\bibinfo {year}
  {2016})}\BibitemShut {NoStop}%
\bibitem [{\citenamefont {Birnstiel}(2024)}]{BirnstielReview2024}%
  \BibitemOpen
  \bibfield  {author} {\bibinfo {author} {\bibfnamefont {T.}~\bibnamefont
  {Birnstiel}},\ }\bibfield  {title} {\bibinfo {title} {Dust growth and
  evolution in protoplanetary disks},\ }\href
  {https://doi.org/https://doi.org/10.1146/annurev-astro-071221-052705}
  {\bibfield  {journal} {\bibinfo  {journal} {Annual Review of Astronomy and
  Astrophysics}\ }\textbf {\bibinfo {volume} {62}},\ \bibinfo {pages} {157}
  (\bibinfo {year} {2024})}\BibitemShut {NoStop}%
\bibitem [{\citenamefont {Mattsson}\ \emph {et~al.}(2018)\citenamefont
  {Mattsson}, \citenamefont {Bhatnagar}, \citenamefont {Gent},\ and\
  \citenamefont {Villarroel}}]{Mattssonetal2018}%
  \BibitemOpen
  \bibfield  {author} {\bibinfo {author} {\bibfnamefont {L.}~\bibnamefont
  {Mattsson}}, \bibinfo {author} {\bibfnamefont {A.}~\bibnamefont {Bhatnagar}},
  \bibinfo {author} {\bibfnamefont {F.~A.}\ \bibnamefont {Gent}},\ and\
  \bibinfo {author} {\bibfnamefont {B.}~\bibnamefont {Villarroel}},\ }\bibfield
   {title} {\bibinfo {title} {Clustering and dynamic decoupling of dust grains
  in turbulent molecular clouds},\ }\href
  {https://doi.org/10.1093/mnras/sty3369} {\bibfield  {journal} {\bibinfo
  {journal} {Monthly Notices of the Royal Astronomical Society}\ }\textbf
  {\bibinfo {volume} {483}},\ \bibinfo {pages} {5623} (\bibinfo {year}
  {2018})}\BibitemShut {NoStop}%
\bibitem [{\citenamefont {Bec}(2003)}]{Bec2003}%
  \BibitemOpen
  \bibfield  {author} {\bibinfo {author} {\bibfnamefont {J.}~\bibnamefont
  {Bec}},\ }\bibfield  {title} {\bibinfo {title} {Fractal clustering of
  inertial particles in random flows},\ }\href
  {https://doi.org/10.1063/1.1612500} {\bibfield  {journal} {\bibinfo
  {journal} {Physics of Fluids}\ }\textbf {\bibinfo {volume} {15}},\ \bibinfo
  {pages} {L81} (\bibinfo {year} {2003})}\BibitemShut {NoStop}%
\bibitem [{\citenamefont {Balkovsky}\ \emph {et~al.}(2001)\citenamefont
  {Balkovsky}, \citenamefont {Falkovich},\ and\ \citenamefont
  {Fouxon}}]{Balkovskyetal2001}%
  \BibitemOpen
  \bibfield  {author} {\bibinfo {author} {\bibfnamefont {E.}~\bibnamefont
  {Balkovsky}}, \bibinfo {author} {\bibfnamefont {G.}~\bibnamefont
  {Falkovich}},\ and\ \bibinfo {author} {\bibfnamefont {A.}~\bibnamefont
  {Fouxon}},\ }\bibfield  {title} {\bibinfo {title} {Intermittent distribution
  of inertial particles in turbulent flows},\ }\href
  {https://doi.org/10.1103/PhysRevLett.86.2790} {\bibfield  {journal} {\bibinfo
   {journal} {Physical Review Letters}\ }\textbf {\bibinfo {volume} {86}},\
  \bibinfo {pages} {2790} (\bibinfo {year} {2001})}\BibitemShut {NoStop}%
\bibitem [{\citenamefont {Falkovich}\ \emph {et~al.}(2002)\citenamefont
  {Falkovich}, \citenamefont {Fouxon},\ and\ \citenamefont
  {Stepanov}}]{Falkovichetal2002}%
  \BibitemOpen
  \bibfield  {author} {\bibinfo {author} {\bibfnamefont {G.}~\bibnamefont
  {Falkovich}}, \bibinfo {author} {\bibfnamefont {I.}~\bibnamefont {Fouxon}},\
  and\ \bibinfo {author} {\bibfnamefont {M.~G.}\ \bibnamefont {Stepanov}},\
  }\bibfield  {title} {\bibinfo {title} {Acceleration of rain initiation by
  cloud turbulence},\ }\href {https://doi.org/10.1038/nature00983} {\bibfield
  {journal} {\bibinfo  {journal} {Nature}\ }\textbf {\bibinfo {volume} {419}},\
  \bibinfo {pages} {151} (\bibinfo {year} {2002})}\BibitemShut {NoStop}%
\bibitem [{\citenamefont {Wilkinson}\ and\ \citenamefont
  {Mehlig}(2003)}]{WilkinsonMehlig2003}%
  \BibitemOpen
  \bibfield  {author} {\bibinfo {author} {\bibfnamefont {M.}~\bibnamefont
  {Wilkinson}}\ and\ \bibinfo {author} {\bibfnamefont {B.}~\bibnamefont
  {Mehlig}},\ }\bibfield  {title} {\bibinfo {title} {Path coalescence
  transition and its applications},\ }\href
  {https://doi.org/10.1103/PhysRevE.68.040101} {\bibfield  {journal} {\bibinfo
  {journal} {Phys. Rev. E}\ }\textbf {\bibinfo {volume} {68}},\ \bibinfo
  {pages} {040101} (\bibinfo {year} {2003})}\BibitemShut {NoStop}%
\bibitem [{\citenamefont {Wilkinson}\ and\ \citenamefont
  {Mehlig}(2005)}]{WilkinsonMehlig2005}%
  \BibitemOpen
  \bibfield  {author} {\bibinfo {author} {\bibfnamefont {M.}~\bibnamefont
  {Wilkinson}}\ and\ \bibinfo {author} {\bibfnamefont {B.}~\bibnamefont
  {Mehlig}},\ }\bibfield  {title} {\bibinfo {title} {Caustics in turbulent
  aerosols},\ }\href {https://doi.org/10.1209/epl/i2004-10532-7} {\bibfield
  {journal} {\bibinfo  {journal} {Europhysics Letters}\ }\textbf {\bibinfo
  {volume} {71}},\ \bibinfo {pages} {186} (\bibinfo {year} {2005})}\BibitemShut
  {NoStop}%
\bibitem [{\citenamefont {Bec}\ \emph {et~al.}(2007)\citenamefont {Bec},
  \citenamefont {Biferale}, \citenamefont {Cencini}, \citenamefont {Lanotte},\
  and\ \citenamefont {Toschi}}]{Becetal2007}%
  \BibitemOpen
  \bibfield  {author} {\bibinfo {author} {\bibfnamefont {J.}~\bibnamefont
  {Bec}}, \bibinfo {author} {\bibfnamefont {L.}~\bibnamefont {Biferale}},
  \bibinfo {author} {\bibfnamefont {M.}~\bibnamefont {Cencini}}, \bibinfo
  {author} {\bibfnamefont {A.~S.}\ \bibnamefont {Lanotte}},\ and\ \bibinfo
  {author} {\bibfnamefont {F.}~\bibnamefont {Toschi}},\ }\bibfield  {title}
  {\bibinfo {title} {Heavy particle concentration in turbulence at dissipative
  and inertial scales},\ }\href {https://doi.org/10.1103/PhysRevLett.98.084502}
  {\bibfield  {journal} {\bibinfo  {journal} {Physical Review Letters}\
  }\textbf {\bibinfo {volume} {98}},\ \bibinfo {pages} {084502} (\bibinfo
  {year} {2007})}\BibitemShut {NoStop}%
\bibitem [{\citenamefont {Gustavsson}\ and\ \citenamefont
  {Mehlig}(2011)}]{GustavssonMehlig2009}%
  \BibitemOpen
  \bibfield  {author} {\bibinfo {author} {\bibfnamefont {K.}~\bibnamefont
  {Gustavsson}}\ and\ \bibinfo {author} {\bibfnamefont {B.}~\bibnamefont
  {Mehlig}},\ }\bibfield  {title} {\bibinfo {title} {Statistical model for
  collisions and clustering of inertial particles in turbulence},\ }\href
  {https://doi.org/10.1209/0295-5075/96/60012} {\bibfield  {journal} {\bibinfo
  {journal} {Europhysics Letters}\ }\textbf {\bibinfo {volume} {96}},\ \bibinfo
  {pages} {60012} (\bibinfo {year} {2011})}\BibitemShut {NoStop}%
\bibitem [{\citenamefont {Bec}\ \emph {et~al.}(2014)\citenamefont {Bec},
  \citenamefont {Homann},\ and\ \citenamefont {Ray}}]{BecHomannRay2014}%
  \BibitemOpen
  \bibfield  {author} {\bibinfo {author} {\bibfnamefont {J.}~\bibnamefont
  {Bec}}, \bibinfo {author} {\bibfnamefont {H.}~\bibnamefont {Homann}},\ and\
  \bibinfo {author} {\bibfnamefont {S.~S.}\ \bibnamefont {Ray}},\ }\bibfield
  {title} {\bibinfo {title} {Gravity-driven enhancement of heavy particle
  clustering in turbulent flow},\ }\href
  {https://doi.org/10.1103/PhysRevLett.112.184501} {\bibfield  {journal}
  {\bibinfo  {journal} {Phys. Rev. Lett.}\ }\textbf {\bibinfo {volume} {112}},\
  \bibinfo {pages} {184501} (\bibinfo {year} {2014})}\BibitemShut {NoStop}%
\bibitem [{\citenamefont {Saw}\ \emph {et~al.}(2014)\citenamefont {Saw},
  \citenamefont {Bewley}, \citenamefont {Bodenschatz}, \citenamefont
  {Sankar~Ray},\ and\ \citenamefont {Bec}}]{Sawetal2014}%
  \BibitemOpen
  \bibfield  {author} {\bibinfo {author} {\bibfnamefont {E.-W.}\ \bibnamefont
  {Saw}}, \bibinfo {author} {\bibfnamefont {G.~P.}\ \bibnamefont {Bewley}},
  \bibinfo {author} {\bibfnamefont {E.}~\bibnamefont {Bodenschatz}}, \bibinfo
  {author} {\bibfnamefont {S.}~\bibnamefont {Sankar~Ray}},\ and\ \bibinfo
  {author} {\bibfnamefont {J.}~\bibnamefont {Bec}},\ }\bibfield  {title}
  {\bibinfo {title} {Extreme fluctuations of the relative velocities between
  droplets in turbulent airflow},\ }\href {https://doi.org/10.1063/1.4900848}
  {\bibfield  {journal} {\bibinfo  {journal} {Physics of Fluids}\ }\textbf
  {\bibinfo {volume} {26}},\ \bibinfo {pages} {111702} (\bibinfo {year}
  {2014})}\BibitemShut {NoStop}%
\bibitem [{\citenamefont {Bec}\ \emph {et~al.}(2016)\citenamefont {Bec},
  \citenamefont {Ray}, \citenamefont {Saw},\ and\ \citenamefont
  {Homann}}]{Becetal2016}%
  \BibitemOpen
  \bibfield  {author} {\bibinfo {author} {\bibfnamefont {J.}~\bibnamefont
  {Bec}}, \bibinfo {author} {\bibfnamefont {S.~S.}\ \bibnamefont {Ray}},
  \bibinfo {author} {\bibfnamefont {E.~W.}\ \bibnamefont {Saw}},\ and\ \bibinfo
  {author} {\bibfnamefont {H.}~\bibnamefont {Homann}},\ }\bibfield  {title}
  {\bibinfo {title} {Abrupt growth of large aggregates by correlated
  coalescences in turbulent flow},\ }\href
  {https://doi.org/10.1103/PhysRevE.93.031102} {\bibfield  {journal} {\bibinfo
  {journal} {Phys. Rev. E}\ }\textbf {\bibinfo {volume} {93}},\ \bibinfo
  {pages} {031102} (\bibinfo {year} {2016})}\BibitemShut {NoStop}%
\bibitem [{\citenamefont {James}\ and\ \citenamefont
  {Ray}(2017)}]{JamesRay2017}%
  \BibitemOpen
  \bibfield  {author} {\bibinfo {author} {\bibfnamefont {M.}~\bibnamefont
  {James}}\ and\ \bibinfo {author} {\bibfnamefont {S.~S.}\ \bibnamefont
  {Ray}},\ }\bibfield  {title} {\bibinfo {title} {Enhanced droplet collision
  rates and impact velocities in turbulent flows: The effect of poly-dispersity
  and transient phases},\ }\href {https://doi.org/10.1038/s41598-017-12093-0}
  {\bibfield  {journal} {\bibinfo  {journal} {Scientific Reports}\ }\textbf
  {\bibinfo {volume} {7}},\ \bibinfo {pages} {12231} (\bibinfo {year}
  {2017})}\BibitemShut {NoStop}%
\bibitem [{\citenamefont {Ray}(2018)}]{Ray2018}%
  \BibitemOpen
  \bibfield  {author} {\bibinfo {author} {\bibfnamefont {S.~S.}\ \bibnamefont
  {Ray}},\ }\bibfield  {title} {\bibinfo {title} {Non-intermittent turbulence:
  Lagrangian chaos and irreversibility},\ }\href
  {https://doi.org/10.1103/PhysRevFluids.3.072601} {\bibfield  {journal}
  {\bibinfo  {journal} {Phys. Rev. Fluids}\ }\textbf {\bibinfo {volume} {3}},\
  \bibinfo {pages} {072601} (\bibinfo {year} {2018})}\BibitemShut {NoStop}%
\bibitem [{\citenamefont {Picardo}\ \emph {et~al.}(2019)\citenamefont
  {Picardo}, \citenamefont {Agasthya}, \citenamefont {Govindarajan},\ and\
  \citenamefont {Ray}}]{Picardoetal2019}%
  \BibitemOpen
  \bibfield  {author} {\bibinfo {author} {\bibfnamefont {J.~R.}\ \bibnamefont
  {Picardo}}, \bibinfo {author} {\bibfnamefont {L.}~\bibnamefont {Agasthya}},
  \bibinfo {author} {\bibfnamefont {R.}~\bibnamefont {Govindarajan}},\ and\
  \bibinfo {author} {\bibfnamefont {S.~S.}\ \bibnamefont {Ray}},\ }\bibfield
  {title} {\bibinfo {title} {Flow structures govern particle collisions in
  turbulence},\ }\href {https://doi.org/10.1103/PhysRevFluids.4.032601}
  {\bibfield  {journal} {\bibinfo  {journal} {Phys. Rev. Fluids}\ }\textbf
  {\bibinfo {volume} {4}},\ \bibinfo {pages} {032601} (\bibinfo {year}
  {2019})}\BibitemShut {NoStop}%
\bibitem [{\citenamefont {Yang}\ \emph {et~al.}(2014)\citenamefont {Yang},
  \citenamefont {Wang}, \citenamefont {Shi}, \citenamefont {Xiao},
  \citenamefont {He},\ and\ \citenamefont {Chen}}]{Yangetal2014}%
  \BibitemOpen
  \bibfield  {author} {\bibinfo {author} {\bibfnamefont {Y.}~\bibnamefont
  {Yang}}, \bibinfo {author} {\bibfnamefont {J.}~\bibnamefont {Wang}}, \bibinfo
  {author} {\bibfnamefont {Y.}~\bibnamefont {Shi}}, \bibinfo {author}
  {\bibfnamefont {Z.}~\bibnamefont {Xiao}}, \bibinfo {author} {\bibfnamefont
  {X.~T.}\ \bibnamefont {He}},\ and\ \bibinfo {author} {\bibfnamefont
  {S.}~\bibnamefont {Chen}},\ }\bibfield  {title} {\bibinfo {title}
  {Interactions between inertial particles and shocklets in compressible
  turbulent flow},\ }\href {https://doi.org/10.1063/1.4896267} {\bibfield
  {journal} {\bibinfo  {journal} {Physics of Fluids}\ }\textbf {\bibinfo
  {volume} {26}},\ \bibinfo {pages} {091702} (\bibinfo {year}
  {2014})}\BibitemShut {NoStop}%
\bibitem [{\citenamefont {Xia}\ \emph {et~al.}(2016)\citenamefont {Xia},
  \citenamefont {Shi}, \citenamefont {Zhang},\ and\ \citenamefont
  {Chen}}]{Xiaetal2016}%
  \BibitemOpen
  \bibfield  {author} {\bibinfo {author} {\bibfnamefont {Z.}~\bibnamefont
  {Xia}}, \bibinfo {author} {\bibfnamefont {Y.}~\bibnamefont {Shi}}, \bibinfo
  {author} {\bibfnamefont {Q.}~\bibnamefont {Zhang}},\ and\ \bibinfo {author}
  {\bibfnamefont {S.}~\bibnamefont {Chen}},\ }\bibfield  {title} {\bibinfo
  {title} {Modulation to compressible homogenous turbulence by heavy point
  particles. i. effect of particles’ density},\ }\href
  {https://doi.org/10.1063/1.4939794} {\bibfield  {journal} {\bibinfo
  {journal} {Physics of Fluids}\ }\textbf {\bibinfo {volume} {28}},\ \bibinfo
  {pages} {016103} (\bibinfo {year} {2016})}\BibitemShut {NoStop}%
\bibitem [{\citenamefont {Mattsson}\ \emph {et~al.}(2019)\citenamefont
  {Mattsson}, \citenamefont {Fynbo},\ and\ \citenamefont
  {Villarroel}}]{Mattssonetal2019}%
  \BibitemOpen
  \bibfield  {author} {\bibinfo {author} {\bibfnamefont {L.}~\bibnamefont
  {Mattsson}}, \bibinfo {author} {\bibfnamefont {J.~P.~U.}\ \bibnamefont
  {Fynbo}},\ and\ \bibinfo {author} {\bibfnamefont {B.}~\bibnamefont
  {Villarroel}},\ }\bibfield  {title} {\bibinfo {title} {Small-scale clustering
  of nano-dust grains in supersonic turbulence},\ }\href
  {https://doi.org/10.1093/mnras/stz2957} {\bibfield  {journal} {\bibinfo
  {journal} {Monthly Notices of the Royal Astronomical Society}\ }\textbf
  {\bibinfo {volume} {490}},\ \bibinfo {pages} {5788} (\bibinfo {year}
  {2019})},\ \Eprint
  {https://arxiv.org/abs/https://academic.oup.com/mnras/article-pdf/490/4/5788/30820647/stz2957.pdf}
  {https://academic.oup.com/mnras/article-pdf/490/4/5788/30820647/stz2957.pdf}
  \BibitemShut {NoStop}%
\bibitem [{\citenamefont {Mattsson}(2020)}]{Mattsson2020}%
  \BibitemOpen
  \bibfield  {author} {\bibinfo {author} {\bibfnamefont {L.}~\bibnamefont
  {Mattsson}},\ }\bibfield  {title} {\bibinfo {title} {On the grain-sized
  distribution of turbulent dust growth},\ }\href
  {https://doi.org/10.1093/mnras/staa3114} {\bibfield  {journal} {\bibinfo
  {journal} {Monthly Notices of the Royal Astronomical Society}\ }\textbf
  {\bibinfo {volume} {499}},\ \bibinfo {pages} {6035} (\bibinfo {year}
  {2020})}\BibitemShut {NoStop}%
\bibitem [{\citenamefont {Gerosa}\ \emph {et~al.}(2023)\citenamefont {Gerosa},
  \citenamefont {M{\'e}heut},\ and\ \citenamefont {Bec}}]{Gerosaetal2023}%
  \BibitemOpen
  \bibfield  {author} {\bibinfo {author} {\bibfnamefont {F.~A.}\ \bibnamefont
  {Gerosa}}, \bibinfo {author} {\bibfnamefont {H.}~\bibnamefont {M{\'e}heut}},\
  and\ \bibinfo {author} {\bibfnamefont {J.}~\bibnamefont {Bec}},\ }\bibfield
  {title} {\bibinfo {title} {Clusters of heavy particles in two-dimensional
  keplerian turbulence},\ }\href
  {https://doi.org/10.1140/epjp/s13360-022-03585-8} {\bibfield  {journal}
  {\bibinfo  {journal} {The European Physical Journal Plus}\ }\textbf {\bibinfo
  {volume} {138}},\ \bibinfo {pages} {9} (\bibinfo {year} {2023})}\BibitemShut
  {NoStop}%
\bibitem [{\citenamefont {Eaton}\ and\ \citenamefont
  {Fessler}(1994)}]{EatonFessler1994}%
  \BibitemOpen
  \bibfield  {author} {\bibinfo {author} {\bibfnamefont {J.}~\bibnamefont
  {Eaton}}\ and\ \bibinfo {author} {\bibfnamefont {J.}~\bibnamefont
  {Fessler}},\ }\bibfield  {title} {\bibinfo {title} {Preferential
  concentration of particles by turbulence},\ }\href
  {https://doi.org/https://doi.org/10.1016/0301-9322(94)90072-8} {\bibfield
  {journal} {\bibinfo  {journal} {International Journal of Multiphase Flow}\
  }\textbf {\bibinfo {volume} {20}},\ \bibinfo {pages} {169} (\bibinfo {year}
  {1994})}\BibitemShut {NoStop}%
\bibitem [{\citenamefont {{Dominik}}\ \emph {et~al.}(2007)\citenamefont
  {{Dominik}}, \citenamefont {{Blum}}, \citenamefont {{Cuzzi}},\ and\
  \citenamefont {{Wurm}}}]{Dominiketal2007}%
  \BibitemOpen
  \bibfield  {author} {\bibinfo {author} {\bibfnamefont {C.}~\bibnamefont
  {{Dominik}}}, \bibinfo {author} {\bibfnamefont {J.}~\bibnamefont {{Blum}}},
  \bibinfo {author} {\bibfnamefont {J.~N.}\ \bibnamefont {{Cuzzi}}},\ and\
  \bibinfo {author} {\bibfnamefont {G.}~\bibnamefont {{Wurm}}},\ }\bibfield
  {title} {\bibinfo {title} {{Growth of Dust as the Initial Step Toward Planet
  Formation}},\ }in\ \href {https://doi.org/10.48550/arXiv.astro-ph/0602617}
  {\emph {\bibinfo {booktitle} {Protostars and Planets V}}},\ \bibinfo {editor}
  {edited by\ \bibinfo {editor} {\bibfnamefont {B.}~\bibnamefont {{Reipurth}}},
  \bibinfo {editor} {\bibfnamefont {D.}~\bibnamefont {{Jewitt}}},\ and\
  \bibinfo {editor} {\bibfnamefont {K.}~\bibnamefont {{Keil}}}}\ (\bibinfo
  {year} {2007})\ p.\ \bibinfo {pages} {783},\ \Eprint
  {https://arxiv.org/abs/astro-ph/0602617} {arXiv:astro-ph/0602617 [astro-ph]}
  \BibitemShut {NoStop}%
\bibitem [{\citenamefont {Johansen}\ \emph {et~al.}(2007)\citenamefont
  {Johansen}, \citenamefont {Oishi}, \citenamefont {Low}, \citenamefont
  {Klahr}, \citenamefont {Henning},\ and\ \citenamefont
  {Youdin}}]{Johansenetal2007}%
  \BibitemOpen
  \bibfield  {author} {\bibinfo {author} {\bibfnamefont {A.}~\bibnamefont
  {Johansen}}, \bibinfo {author} {\bibfnamefont {J.~S.}\ \bibnamefont {Oishi}},
  \bibinfo {author} {\bibfnamefont {M.-M.~M.}\ \bibnamefont {Low}}, \bibinfo
  {author} {\bibfnamefont {H.}~\bibnamefont {Klahr}}, \bibinfo {author}
  {\bibfnamefont {T.}~\bibnamefont {Henning}},\ and\ \bibinfo {author}
  {\bibfnamefont {A.}~\bibnamefont {Youdin}},\ }\bibfield  {title} {\bibinfo
  {title} {Rapid planetesimal formation in turbulent circumstellar disks},\
  }\href {https://doi.org/10.1038/nature06086} {\bibfield  {journal} {\bibinfo
  {journal} {Nature}\ }\textbf {\bibinfo {volume} {448}},\ \bibinfo {pages}
  {1022} (\bibinfo {year} {2007})}\BibitemShut {NoStop}%
\bibitem [{\citenamefont {Wilkinson}\ \emph {et~al.}(2008)\citenamefont
  {Wilkinson}, \citenamefont {Mehlig},\ and\ \citenamefont
  {Uski}}]{Wilkinsonetal2008}%
  \BibitemOpen
  \bibfield  {author} {\bibinfo {author} {\bibfnamefont {M.}~\bibnamefont
  {Wilkinson}}, \bibinfo {author} {\bibfnamefont {B.}~\bibnamefont {Mehlig}},\
  and\ \bibinfo {author} {\bibfnamefont {V.}~\bibnamefont {Uski}},\ }\bibfield
  {title} {\bibinfo {title} {Stokes trapping and planet formation},\ }\href
  {https://doi.org/10.1086/533533} {\bibfield  {journal} {\bibinfo  {journal}
  {The Astrophysical Journal Supplement Series}\ }\textbf {\bibinfo {volume}
  {176}},\ \bibinfo {pages} {484} (\bibinfo {year} {2008})}\BibitemShut
  {NoStop}%
\bibitem [{\citenamefont {Cuzzi}\ \emph {et~al.}(2008)\citenamefont {Cuzzi},
  \citenamefont {Hogan},\ and\ \citenamefont {Shariff}}]{Cuzzietal2008}%
  \BibitemOpen
  \bibfield  {author} {\bibinfo {author} {\bibfnamefont {J.~N.}\ \bibnamefont
  {Cuzzi}}, \bibinfo {author} {\bibfnamefont {R.~C.}\ \bibnamefont {Hogan}},\
  and\ \bibinfo {author} {\bibfnamefont {K.}~\bibnamefont {Shariff}},\
  }\bibfield  {title} {\bibinfo {title} {Toward planetesimals: Dense chondrule
  clumps in the protoplanetary nebula},\ }\href
  {https://doi.org/10.1086/591239} {\bibfield  {journal} {\bibinfo  {journal}
  {The Astrophysical Journal}\ }\textbf {\bibinfo {volume} {687}},\ \bibinfo
  {pages} {1432} (\bibinfo {year} {2008})}\BibitemShut {NoStop}%
\bibitem [{\citenamefont {Youdin}\ and\ \citenamefont
  {Goodman}(2005)}]{YoudinGoodman2005}%
  \BibitemOpen
  \bibfield  {author} {\bibinfo {author} {\bibfnamefont {A.~N.}\ \bibnamefont
  {Youdin}}\ and\ \bibinfo {author} {\bibfnamefont {J.}~\bibnamefont
  {Goodman}},\ }\bibfield  {title} {\bibinfo {title} {Streaming instabilities
  in protoplanetary disks},\ }\href {https://doi.org/10.1086/426895} {\bibfield
   {journal} {\bibinfo  {journal} {The Astrophysical Journal}\ }\textbf
  {\bibinfo {volume} {620}},\ \bibinfo {pages} {459} (\bibinfo {year}
  {2005})}\BibitemShut {NoStop}%
\bibitem [{\citenamefont {Barge}\ and\ \citenamefont
  {Sommeria}(1995)}]{BargeSommeria1995}%
  \BibitemOpen
  \bibfield  {author} {\bibinfo {author} {\bibfnamefont {P.}~\bibnamefont
  {Barge}}\ and\ \bibinfo {author} {\bibfnamefont {J.}~\bibnamefont
  {Sommeria}},\ }\href {https://arxiv.org/abs/astro-ph/9501050} {\bibinfo
  {title} {Did planet formation begin inside persistent gaseous vortices?}}
  (\bibinfo {year} {1995}),\ \Eprint {https://arxiv.org/abs/astro-ph/9501050}
  {arXiv:astro-ph/9501050 [astro-ph]} \BibitemShut {NoStop}%
\bibitem [{\citenamefont {Heng}\ and\ \citenamefont
  {Kenyon}(2010)}]{HengKenyon2010}%
  \BibitemOpen
  \bibfield  {author} {\bibinfo {author} {\bibfnamefont {K.}~\bibnamefont
  {Heng}}\ and\ \bibinfo {author} {\bibfnamefont {S.~J.}\ \bibnamefont
  {Kenyon}},\ }\bibfield  {title} {\bibinfo {title} {Vortices as nurseries for
  planetesimal formation in protoplanetary discs},\ }\href
  {https://doi.org/10.1111/j.1365-2966.2010.17208.x} {\bibfield  {journal}
  {\bibinfo  {journal} {Monthly Notices of the Royal Astronomical Society}\
  }\textbf {\bibinfo {volume} {408}},\ \bibinfo {pages} {1476} (\bibinfo {year}
  {2010})},\ \Eprint
  {https://arxiv.org/abs/https://academic.oup.com/mnras/article-pdf/408/3/1476/18578673/mnras0408-1476.pdf}
  {https://academic.oup.com/mnras/article-pdf/408/3/1476/18578673/mnras0408-1476.pdf}
  \BibitemShut {NoStop}%
\bibitem [{\citenamefont {Lyra}\ and\ \citenamefont {Lin}(2013)}]{LyraLin2013}%
  \BibitemOpen
  \bibfield  {author} {\bibinfo {author} {\bibfnamefont {W.}~\bibnamefont
  {Lyra}}\ and\ \bibinfo {author} {\bibfnamefont {M.-K.}\ \bibnamefont {Lin}},\
  }\bibfield  {title} {\bibinfo {title} {Steady state dust distributions in
  disk vortices: Observational predictions and applications to transitional
  disks},\ }\href {https://doi.org/10.1088/0004-637X/775/1/17} {\bibfield
  {journal} {\bibinfo  {journal} {The Astrophysical Journal}\ }\textbf
  {\bibinfo {volume} {775}},\ \bibinfo {pages} {17} (\bibinfo {year}
  {2013})}\BibitemShut {NoStop}%
\bibitem [{\citenamefont {Gibbons}\ \emph {et~al.}(2015)\citenamefont
  {Gibbons}, \citenamefont {Mamatsashvili},\ and\ \citenamefont
  {Rice}}]{Gibbonetal2015}%
  \BibitemOpen
  \bibfield  {author} {\bibinfo {author} {\bibfnamefont {P.~G.}\ \bibnamefont
  {Gibbons}}, \bibinfo {author} {\bibfnamefont {G.~R.}\ \bibnamefont
  {Mamatsashvili}},\ and\ \bibinfo {author} {\bibfnamefont {W.~K.~M.}\
  \bibnamefont {Rice}},\ }\bibfield  {title} {\bibinfo {title} {Planetesimal
  formation in self-gravitating discs – dust trapping by vortices},\ }\href
  {https://doi.org/10.1093/mnras/stv1766} {\bibfield  {journal} {\bibinfo
  {journal} {Monthly Notices of the Royal Astronomical Society}\ }\textbf
  {\bibinfo {volume} {453}},\ \bibinfo {pages} {4232} (\bibinfo {year}
  {2015})},\ \Eprint
  {https://arxiv.org/abs/https://academic.oup.com/mnras/article-pdf/453/4/4232/8033896/stv1766.pdf}
  {https://academic.oup.com/mnras/article-pdf/453/4/4232/8033896/stv1766.pdf}
  \BibitemShut {NoStop}%
\bibitem [{\citenamefont {{Whipple}}(1972)}]{Whipple1972}%
  \BibitemOpen
  \bibfield  {author} {\bibinfo {author} {\bibfnamefont {F.~L.}\ \bibnamefont
  {{Whipple}}},\ }\bibfield  {title} {\bibinfo {title} {{On certain aerodynamic
  processes for asteroids and comets}},\ }in\ \href@noop {} {\emph {\bibinfo
  {booktitle} {From Plasma to Planet}}},\ \bibinfo {editor} {edited by\
  \bibinfo {editor} {\bibfnamefont {A.}~\bibnamefont {{Elvius}}}}\ (\bibinfo
  {year} {1972})\ p.\ \bibinfo {pages} {211}\BibitemShut {NoStop}%
\bibitem [{\citenamefont {{Pinilla, P.}}\ \emph {et~al.}(2012)\citenamefont
  {{Pinilla, P.}}, \citenamefont {{Benisty, M.}},\ and\ \citenamefont
  {{Birnstiel, T.}}}]{Pinilla2012}%
  \BibitemOpen
  \bibfield  {author} {\bibinfo {author} {\bibnamefont {{Pinilla, P.}}},
  \bibinfo {author} {\bibnamefont {{Benisty, M.}}},\ and\ \bibinfo {author}
  {\bibnamefont {{Birnstiel, T.}}},\ }\bibfield  {title} {\bibinfo {title}
  {Ring shaped dust accumulation in transition disks},\ }\href
  {https://doi.org/10.1051/0004-6361/201219315} {\bibfield  {journal} {\bibinfo
   {journal} {A\&A}\ }\textbf {\bibinfo {volume} {545}},\ \bibinfo {pages}
  {A81} (\bibinfo {year} {2012})}\BibitemShut {NoStop}%
\bibitem [{\citenamefont {{Youdin}}\ and\ \citenamefont
  {{Shu}}(2002)}]{YoudinShu2002}%
  \BibitemOpen
  \bibfield  {author} {\bibinfo {author} {\bibfnamefont {A.~N.}\ \bibnamefont
  {{Youdin}}}\ and\ \bibinfo {author} {\bibfnamefont {F.~H.}\ \bibnamefont
  {{Shu}}},\ }\bibfield  {title} {\bibinfo {title} {{Planetesimal Formation by
  Gravitational Instability}},\ }\href {https://doi.org/10.1086/343109}
  {\bibfield  {journal} {\bibinfo  {journal} {\apj}\ }\textbf {\bibinfo
  {volume} {580}},\ \bibinfo {pages} {494} (\bibinfo {year} {2002})},\ \Eprint
  {https://arxiv.org/abs/astro-ph/0207536} {arXiv:astro-ph/0207536 [astro-ph]}
  \BibitemShut {NoStop}%
\bibitem [{\citenamefont {Rice}\ \emph {et~al.}(2004)\citenamefont {Rice},
  \citenamefont {Lodato}, \citenamefont {Pringle}, \citenamefont {Armitage},\
  and\ \citenamefont {Bonnell}}]{Riceetal2004}%
  \BibitemOpen
  \bibfield  {author} {\bibinfo {author} {\bibfnamefont {W.~K.~M.}\
  \bibnamefont {Rice}}, \bibinfo {author} {\bibfnamefont {G.}~\bibnamefont
  {Lodato}}, \bibinfo {author} {\bibfnamefont {J.~E.}\ \bibnamefont {Pringle}},
  \bibinfo {author} {\bibfnamefont {P.~J.}\ \bibnamefont {Armitage}},\ and\
  \bibinfo {author} {\bibfnamefont {I.~A.}\ \bibnamefont {Bonnell}},\
  }\bibfield  {title} {\bibinfo {title} {Accelerated planetesimal growth in
  self-gravitating protoplanetary discs},\ }\href
  {https://doi.org/10.1111/j.1365-2966.2004.08339.x} {\bibfield  {journal}
  {\bibinfo  {journal} {Monthly Notices of the Royal Astronomical Society}\
  }\textbf {\bibinfo {volume} {355}},\ \bibinfo {pages} {543} (\bibinfo {year}
  {2004})},\ \Eprint
  {https://arxiv.org/abs/https://academic.oup.com/mnras/article-pdf/355/2/543/2918959/355-2-543.pdf}
  {https://academic.oup.com/mnras/article-pdf/355/2/543/2918959/355-2-543.pdf}
  \BibitemShut {NoStop}%
\bibitem [{\citenamefont {Gibbons}\ \emph {et~al.}(2012)\citenamefont
  {Gibbons}, \citenamefont {Rice},\ and\ \citenamefont
  {Mamatsashvili}}]{Gibbonsetal2012}%
  \BibitemOpen
  \bibfield  {author} {\bibinfo {author} {\bibfnamefont {P.~G.}\ \bibnamefont
  {Gibbons}}, \bibinfo {author} {\bibfnamefont {W.~K.~M.}\ \bibnamefont
  {Rice}},\ and\ \bibinfo {author} {\bibfnamefont {G.~R.}\ \bibnamefont
  {Mamatsashvili}},\ }\bibfield  {title} {\bibinfo {title} {Planetesimal
  formation in self-gravitating discs},\ }\href
  {https://doi.org/10.1111/j.1365-2966.2012.21731.x} {\bibfield  {journal}
  {\bibinfo  {journal} {Monthly Notices of the Royal Astronomical Society}\
  }\textbf {\bibinfo {volume} {426}},\ \bibinfo {pages} {1444} (\bibinfo {year}
  {2012})},\ \Eprint
  {https://arxiv.org/abs/https://academic.oup.com/mnras/article-pdf/426/2/1444/2974242/426-2-1444.pdf}
  {https://academic.oup.com/mnras/article-pdf/426/2/1444/2974242/426-2-1444.pdf}
  \BibitemShut {NoStop}%
\bibitem [{\citenamefont {Burgers}(1948)}]{Burgers1948}%
  \BibitemOpen
  \bibfield  {author} {\bibinfo {author} {\bibfnamefont {J.}~\bibnamefont
  {Burgers}},\ }\bibfield  {title} {\bibinfo {title} {A mathematical model
  illustrating the theory of turbulence},\ }\href
  {https://doi.org/https://doi.org/10.1016/S0065-2156(08)70100-5} {\bibfield
  {journal} {\bibinfo  {journal} {Elsevier}\ }\bibinfo {series} {Advances in
  Applied Mechanics},\ \textbf {\bibinfo {volume} {1}},\ \bibinfo {pages} {171}
  (\bibinfo {year} {1948})}\BibitemShut {NoStop}%
\bibitem [{\citenamefont {Frisch}\ and\ \citenamefont
  {Bec}(2001)}]{FrischBecReview2001}%
  \BibitemOpen
  \bibfield  {author} {\bibinfo {author} {\bibfnamefont {U.}~\bibnamefont
  {Frisch}}\ and\ \bibinfo {author} {\bibfnamefont {J.}~\bibnamefont {Bec}},\
  }\bibinfo {title} {Burgulence},\ in\ \href
  {https://doi.org/10.1007/3-540-45674-0_7} {\emph {\bibinfo {booktitle} {New
  trends in turbulence Turbulence: nouveaux aspects: 31 July -- 1 September
  2000}}},\ \bibinfo {editor} {edited by\ \bibinfo {editor} {\bibfnamefont
  {M.}~\bibnamefont {Lesieur}}, \bibinfo {editor} {\bibfnamefont
  {A.}~\bibnamefont {Yaglom}},\ and\ \bibinfo {editor} {\bibfnamefont
  {F.}~\bibnamefont {David}}}\ (\bibinfo  {publisher} {Springer Berlin
  Heidelberg},\ \bibinfo {address} {Berlin, Heidelberg},\ \bibinfo {year}
  {2001})\ pp.\ \bibinfo {pages} {341--383}\BibitemShut {NoStop}%
\bibitem [{\citenamefont {Bec}\ and\ \citenamefont
  {Khanin}(2007)}]{BecKhaninReview2007}%
  \BibitemOpen
  \bibfield  {author} {\bibinfo {author} {\bibfnamefont {J.}~\bibnamefont
  {Bec}}\ and\ \bibinfo {author} {\bibfnamefont {K.}~\bibnamefont {Khanin}},\
  }\bibfield  {title} {\bibinfo {title} {Burgers turbulence},\ }\href
  {https://doi.org/https://doi.org/10.1016/j.physrep.2007.04.002} {\bibfield
  {journal} {\bibinfo  {journal} {Physics Reports}\ }\textbf {\bibinfo {volume}
  {447}},\ \bibinfo {pages} {1} (\bibinfo {year} {2007})}\BibitemShut {NoStop}%
\bibitem [{\citenamefont {Chekhlov}\ and\ \citenamefont
  {Yakhot}(1995)}]{ChekhlovYakhot1995}%
  \BibitemOpen
  \bibfield  {author} {\bibinfo {author} {\bibfnamefont {A.}~\bibnamefont
  {Chekhlov}}\ and\ \bibinfo {author} {\bibfnamefont {V.}~\bibnamefont
  {Yakhot}},\ }\bibfield  {title} {\bibinfo {title} {Kolmogorov turbulence in a
  random-force-driven burgers equation},\ }\href
  {https://doi.org/10.1103/PhysRevE.51.R2739} {\bibfield  {journal} {\bibinfo
  {journal} {Phys. Rev. E}\ }\textbf {\bibinfo {volume} {51}},\ \bibinfo
  {pages} {R2739} (\bibinfo {year} {1995})}\BibitemShut {NoStop}%
\bibitem [{\citenamefont {Hayot}\ and\ \citenamefont
  {Jayaprakash}(1997)}]{HayotJayaprakash1997}%
  \BibitemOpen
  \bibfield  {author} {\bibinfo {author} {\bibfnamefont {F.}~\bibnamefont
  {Hayot}}\ and\ \bibinfo {author} {\bibfnamefont {C.}~\bibnamefont
  {Jayaprakash}},\ }\bibfield  {title} {\bibinfo {title} {From scaling to
  multiscaling in the stochastic burgers equation},\ }\href
  {https://doi.org/10.1103/PhysRevE.56.4259} {\bibfield  {journal} {\bibinfo
  {journal} {Phys. Rev. E}\ }\textbf {\bibinfo {volume} {56}},\ \bibinfo
  {pages} {4259} (\bibinfo {year} {1997})}\BibitemShut {NoStop}%
\bibitem [{\citenamefont {Mitra}\ \emph {et~al.}(2005)\citenamefont {Mitra},
  \citenamefont {Bec}, \citenamefont {Pandit},\ and\ \citenamefont
  {Frisch}}]{Mitraetal2005}%
  \BibitemOpen
  \bibfield  {author} {\bibinfo {author} {\bibfnamefont {D.}~\bibnamefont
  {Mitra}}, \bibinfo {author} {\bibfnamefont {J.}~\bibnamefont {Bec}}, \bibinfo
  {author} {\bibfnamefont {R.}~\bibnamefont {Pandit}},\ and\ \bibinfo {author}
  {\bibfnamefont {U.}~\bibnamefont {Frisch}},\ }\bibfield  {title} {\bibinfo
  {title} {Is multiscaling an artifact in the stochastically forced burgers
  equation?},\ }\href {https://doi.org/10.1103/PhysRevLett.94.194501}
  {\bibfield  {journal} {\bibinfo  {journal} {Phys. Rev. Lett.}\ }\textbf
  {\bibinfo {volume} {94}},\ \bibinfo {pages} {194501} (\bibinfo {year}
  {2005})}\BibitemShut {NoStop}%
\bibitem [{\citenamefont {Frisch}\ \emph {et~al.}(2013)\citenamefont {Frisch},
  \citenamefont {Ray}, \citenamefont {Sahoo}, \citenamefont {Banerjee},\ and\
  \citenamefont {Pandit}}]{Frischetal2013}%
  \BibitemOpen
  \bibfield  {author} {\bibinfo {author} {\bibfnamefont {U.}~\bibnamefont
  {Frisch}}, \bibinfo {author} {\bibfnamefont {S.~S.}\ \bibnamefont {Ray}},
  \bibinfo {author} {\bibfnamefont {G.}~\bibnamefont {Sahoo}}, \bibinfo
  {author} {\bibfnamefont {D.}~\bibnamefont {Banerjee}},\ and\ \bibinfo
  {author} {\bibfnamefont {R.}~\bibnamefont {Pandit}},\ }\bibfield  {title}
  {\bibinfo {title} {Real-space manifestations of bottlenecks in turbulence
  spectra},\ }\href {https://doi.org/10.1103/PhysRevLett.110.064501} {\bibfield
   {journal} {\bibinfo  {journal} {Phys. Rev. Lett.}\ }\textbf {\bibinfo
  {volume} {110}},\ \bibinfo {pages} {064501} (\bibinfo {year}
  {2013})}\BibitemShut {NoStop}%
\bibitem [{\citenamefont {De}\ \emph {et~al.}(2023)\citenamefont {De},
  \citenamefont {Mitra},\ and\ \citenamefont {Pandit}}]{DeMitraPandit2023}%
  \BibitemOpen
  \bibfield  {author} {\bibinfo {author} {\bibfnamefont {S.}~\bibnamefont
  {De}}, \bibinfo {author} {\bibfnamefont {D.}~\bibnamefont {Mitra}},\ and\
  \bibinfo {author} {\bibfnamefont {R.}~\bibnamefont {Pandit}},\ }\bibfield
  {title} {\bibinfo {title} {Dynamic multiscaling in stochastically forced
  burgers turbulence},\ }\href {https://doi.org/10.1038/s41598-023-29056-3}
  {\bibfield  {journal} {\bibinfo  {journal} {Scientific Reports}\ }\textbf
  {\bibinfo {volume} {13}},\ \bibinfo {pages} {7151} (\bibinfo {year}
  {2023})}\BibitemShut {NoStop}%
\bibitem [{\citenamefont {De}\ \emph {et~al.}(2024)\citenamefont {De},
  \citenamefont {Mitra},\ and\ \citenamefont {Pandit}}]{DeMitraPandit2024}%
  \BibitemOpen
  \bibfield  {author} {\bibinfo {author} {\bibfnamefont {S.}~\bibnamefont
  {De}}, \bibinfo {author} {\bibfnamefont {D.}~\bibnamefont {Mitra}},\ and\
  \bibinfo {author} {\bibfnamefont {R.}~\bibnamefont {Pandit}},\ }\bibfield
  {title} {\bibinfo {title} {Uncovering the multifractality of lagrangian pair
  dispersion in shock-dominated turbulence},\ }\href
  {https://doi.org/10.1103/PhysRevResearch.6.L022032} {\bibfield  {journal}
  {\bibinfo  {journal} {Phys. Rev. Res.}\ }\textbf {\bibinfo {volume} {6}},\
  \bibinfo {pages} {L022032} (\bibinfo {year} {2024})}\BibitemShut {NoStop}%
\bibitem [{\citenamefont {Shandarin}\ and\ \citenamefont
  {Zeldovich}(1989)}]{ShandarinZeldovichReview1989}%
  \BibitemOpen
  \bibfield  {author} {\bibinfo {author} {\bibfnamefont {S.~F.}\ \bibnamefont
  {Shandarin}}\ and\ \bibinfo {author} {\bibfnamefont {Y.~B.}\ \bibnamefont
  {Zeldovich}},\ }\bibfield  {title} {\bibinfo {title} {The large-scale
  structure of the universe: Turbulence, intermittency, structures in a
  self-gravitating medium},\ }\href {https://doi.org/10.1103/RevModPhys.61.185}
  {\bibfield  {journal} {\bibinfo  {journal} {Rev. Mod. Phys.}\ }\textbf
  {\bibinfo {volume} {61}},\ \bibinfo {pages} {185} (\bibinfo {year}
  {1989})}\BibitemShut {NoStop}%
\bibitem [{\citenamefont {Matarrese}\ and\ \citenamefont
  {Mohayaee}(2002)}]{MatarreseMohayaee2002}%
  \BibitemOpen
  \bibfield  {author} {\bibinfo {author} {\bibfnamefont {S.}~\bibnamefont
  {Matarrese}}\ and\ \bibinfo {author} {\bibfnamefont {R.}~\bibnamefont
  {Mohayaee}},\ }\bibfield  {title} {\bibinfo {title} {The growth of structure
  in the intergalactic medium},\ }\href
  {https://doi.org/10.1046/j.1365-8711.2002.04944.x} {\bibfield  {journal}
  {\bibinfo  {journal} {Monthly Notices of the Royal Astronomical Society}\
  }\textbf {\bibinfo {volume} {329}},\ \bibinfo {pages} {37} (\bibinfo {year}
  {2002})},\ \Eprint
  {https://arxiv.org/abs/https://academic.oup.com/mnras/article-pdf/329/1/37/3881162/329-1-37.pdf}
  {https://academic.oup.com/mnras/article-pdf/329/1/37/3881162/329-1-37.pdf}
  \BibitemShut {NoStop}%
\bibitem [{\citenamefont {Neate}\ and\ \citenamefont
  {Truman}(2007)}]{NeateTruman2007}%
  \BibitemOpen
  \bibfield  {author} {\bibinfo {author} {\bibfnamefont {A.}~\bibnamefont
  {Neate}}\ and\ \bibinfo {author} {\bibfnamefont {A.}~\bibnamefont {Truman}},\
  }\href {https://arxiv.org/abs/0711.0617} {\bibinfo {title} {On the stochastic
  burgers equation with some applications to turbulence and astrophysics}}
  (\bibinfo {year} {2007}),\ \Eprint {https://arxiv.org/abs/0711.0617}
  {arXiv:0711.0617 [math.PR]} \BibitemShut {NoStop}%
\bibitem [{\citenamefont {Kankaria}(2025)}]{YT-movies}%
  \BibitemOpen
  \bibfield  {author} {\bibinfo {author} {\bibnamefont {Kankaria}},\
  }\href@noop {} {}\bibinfo {howpublished} {See
  \url{https://www.youtube.com/watch?v=FjvfHtVXtzQ} for an animation showing
  the evolution of particle density field for different Stokes numbers.}
  (\bibinfo {year} {2025})\BibitemShut {NoStop}%
\bibitem [{\citenamefont {Cox}\ and\ \citenamefont
  {Matthews}(2002)}]{CoxMathews2002}%
  \BibitemOpen
  \bibfield  {author} {\bibinfo {author} {\bibfnamefont {S.}~\bibnamefont
  {Cox}}\ and\ \bibinfo {author} {\bibfnamefont {P.}~\bibnamefont {Matthews}},\
  }\bibfield  {title} {\bibinfo {title} {Exponential time differencing for
  stiff systems},\ }\href
  {https://doi.org/https://doi.org/10.1006/jcph.2002.6995} {\bibfield
  {journal} {\bibinfo  {journal} {Journal of Computational Physics}\ }\textbf
  {\bibinfo {volume} {176}},\ \bibinfo {pages} {430} (\bibinfo {year}
  {2002})}\BibitemShut {NoStop}%
\bibitem [{\citenamefont {Goldhirsch}(2003)}]{GoldhirschReview2003}%
  \BibitemOpen
  \bibfield  {author} {\bibinfo {author} {\bibfnamefont {I.}~\bibnamefont
  {Goldhirsch}},\ }\bibfield  {title} {\bibinfo {title} {Rapid granular
  flows},\ }\href
  {https://doi.org/https://doi.org/10.1146/annurev.fluid.35.101101.161114}
  {\bibfield  {journal} {\bibinfo  {journal} {Annual Review of Fluid
  Mechanics}\ }\textbf {\bibinfo {volume} {35}},\ \bibinfo {pages} {267}
  (\bibinfo {year} {2003})}\BibitemShut {NoStop}%
\bibitem [{\citenamefont {Bec}\ \emph {et~al.}(2013)\citenamefont {Bec},
  \citenamefont {Musacchio},\ and\ \citenamefont {Ray}}]{Becetal2013}%
  \BibitemOpen
  \bibfield  {author} {\bibinfo {author} {\bibfnamefont {J.}~\bibnamefont
  {Bec}}, \bibinfo {author} {\bibfnamefont {S.}~\bibnamefont {Musacchio}},\
  and\ \bibinfo {author} {\bibfnamefont {S.~S.}\ \bibnamefont {Ray}},\
  }\bibfield  {title} {\bibinfo {title} {Sticky elastic collisions},\ }\href
  {https://doi.org/10.1103/PhysRevE.87.063013} {\bibfield  {journal} {\bibinfo
  {journal} {Phys. Rev. E}\ }\textbf {\bibinfo {volume} {87}},\ \bibinfo
  {pages} {063013} (\bibinfo {year} {2013})}\BibitemShut {NoStop}%
\bibitem [{\citenamefont {Picardo}\ \emph {et~al.}(2020)\citenamefont
  {Picardo}, \citenamefont {Singh}, \citenamefont {Ray},\ and\ \citenamefont
  {Vincenzi}}]{Picardoetal2020}%
  \BibitemOpen
  \bibfield  {author} {\bibinfo {author} {\bibfnamefont {J.~R.}\ \bibnamefont
  {Picardo}}, \bibinfo {author} {\bibfnamefont {R.}~\bibnamefont {Singh}},
  \bibinfo {author} {\bibfnamefont {S.~S.}\ \bibnamefont {Ray}},\ and\ \bibinfo
  {author} {\bibfnamefont {D.}~\bibnamefont {Vincenzi}},\ }\bibfield  {title}
  {\bibinfo {title} {Dynamics of a long chain in turbulent flows: impact of
  vortices},\ }\href {https://doi.org/10.1098/rsta.2019.0405} {\bibfield
  {journal} {\bibinfo  {journal} {Philosophical Transactions of the Royal
  Society A: Mathematical, Physical and Engineering Sciences}\ }\textbf
  {\bibinfo {volume} {378}},\ \bibinfo {pages} {20190405} (\bibinfo {year}
  {2020})},\ \Eprint
  {https://arxiv.org/abs/https://royalsocietypublishing.org/rsta/article-pdf/doi/10.1098/rsta.2019.0405/1317550/rsta.2019.0405.pdf}
  {https://royalsocietypublishing.org/rsta/article-pdf/doi/10.1098/rsta.2019.0405/1317550/rsta.2019.0405.pdf}
  \BibitemShut {NoStop}%
\bibitem [{\citenamefont {Anand}\ \emph {et~al.}(2020)\citenamefont {Anand},
  \citenamefont {Ray},\ and\ \citenamefont {Subramanian}}]{Anandetal2020}%
  \BibitemOpen
  \bibfield  {author} {\bibinfo {author} {\bibfnamefont {P.}~\bibnamefont
  {Anand}}, \bibinfo {author} {\bibfnamefont {S.~S.}\ \bibnamefont {Ray}},\
  and\ \bibinfo {author} {\bibfnamefont {G.}~\bibnamefont {Subramanian}},\
  }\bibfield  {title} {\bibinfo {title} {Orientation dynamics of sedimenting
  anisotropic particles in turbulence},\ }\href
  {https://doi.org/10.1103/PhysRevLett.125.034501} {\bibfield  {journal}
  {\bibinfo  {journal} {Phys. Rev. Lett.}\ }\textbf {\bibinfo {volume} {125}},\
  \bibinfo {pages} {034501} (\bibinfo {year} {2020})}\BibitemShut {NoStop}%
\bibitem [{\citenamefont {Pandey}\ \emph {et~al.}(2019)\citenamefont {Pandey},
  \citenamefont {Perlekar},\ and\ \citenamefont {Mitra}}]{Pandeyetal2019}%
  \BibitemOpen
  \bibfield  {author} {\bibinfo {author} {\bibfnamefont {V.}~\bibnamefont
  {Pandey}}, \bibinfo {author} {\bibfnamefont {P.}~\bibnamefont {Perlekar}},\
  and\ \bibinfo {author} {\bibfnamefont {D.}~\bibnamefont {Mitra}},\ }\bibfield
   {title} {\bibinfo {title} {Clustering and energy spectra in two-dimensional
  dusty gas turbulence},\ }\href {https://doi.org/10.1103/PhysRevE.100.013114}
  {\bibfield  {journal} {\bibinfo  {journal} {Phys. Rev. E}\ }\textbf {\bibinfo
  {volume} {100}},\ \bibinfo {pages} {013114} (\bibinfo {year}
  {2019})}\BibitemShut {NoStop}%
\bibitem [{\citenamefont {Joshi}\ and\ \citenamefont
  {Ray}(2025)}]{JoshiRay2025}%
  \BibitemOpen
  \bibfield  {author} {\bibinfo {author} {\bibfnamefont {H.}~\bibnamefont
  {Joshi}}\ and\ \bibinfo {author} {\bibfnamefont {S.~S.}\ \bibnamefont
  {Ray}},\ }\href {https://arxiv.org/abs/2510.10463} {\bibinfo {title} {The
  significance of two-way coupling in two-dimensional, dusty turbulence}}
  (\bibinfo {year} {2025}),\ \Eprint {https://arxiv.org/abs/2510.10463}
  {arXiv:2510.10463 [physics.flu-dyn]} \BibitemShut {NoStop}%
\end{thebibliography}%



\label{lastpage}
\end{document}